\documentclass[longauth,traditabstract,usenatbib]{aa} 
\usepackage{graphicx}
\usepackage{txfonts}
\usepackage{textcomp}
\usepackage{color}
\usepackage[nointegrals]{wasysym}
\usepackage{amssymb}
\usepackage[separate-uncertainty = true,multi-part-units=single]{siunitx}
\usepackage{natbib}
\usepackage{babel}
\usepackage[normalem]{ulem}
\usepackage[switch]{lineno}

\bibpunct{(}{)}{;}{a}{}{,}

\PassOptionsToPackage{american}{babel}
\PassOptionsToPackage{authoryear}{natbib}

\renewcommand{\deg}{^\circ}

\usepackage[normalem]{ulem}

\usepackage{scrextend}

\begin{document} 

\title{MAGIC observations of the diffuse $\gamma$-ray emission in the vicinity of the Galactic Centre}


%
%
\author{
MAGIC Collaboration \and
V.~A.~Acciari\inst{1} \and
S.~Ansoldi\inst{2,23} \and
L.~A.~Antonelli\inst{3} \and
A.~Arbet Engels\inst{4} \and
D.~Baack\inst{5} \and
A.~Babi\'c\inst{6} \and
B.~Banerjee\inst{7} \and
U.~Barres de Almeida\inst{8} \and
J.~A.~Barrio\inst{9} \and
J.~Becerra Gonz\'alez\inst{1}
% ...
}
\institute { Inst. de Astrof\'isica de Canarias, E-38200 La Laguna, and Universidad de La Laguna, Dpto. Astrof\'isica, E-38206 La Laguna, Tenerife, Spain
\and Universit\`a di Udine, and INFN Trieste, I-33100 Udine, Italy
\and National Institute for Astrophysics (INAF), I-00136 Rome, Italy
\and ETH Zurich, CH-8093 Zurich, Switzerland
\and Technische Universit\"at Dortmund, D-44221 Dortmund, Germany
\and Croatian Consortium: University of Rijeka, Department of Physics, 51000 Rijeka; University of Split - FESB, 21000 Split; University of Zagreb - FER, 10000 Zagreb; University of Osijek, 31000 Osijek; Rudjer Boskovic Institute, 10000 Zagreb, Croatia
\and Saha Institute of Nuclear Physics, HBNI, 1/AF Bidhannagar, Salt Lake, Sector-1, Kolkata 700064, India
\and Centro Brasileiro de Pesquisas F\'isicas (CBPF), 22290-180 URCA, Rio de Janeiro (RJ), Brasil
\and IPARCOS Institute and EMFTEL Department, Universidad Complutense de Madrid, E-28040 Madrid, Spain
\and University of \L\'od\'z, Department of Astrophysics, PL-90236 \L\'od\'z, Poland
\and Universit\`a di Siena and INFN Pisa, I-53100 Siena, Italy
\and Deutsches Elektronen-Synchrotron (DESY), D-15738 Zeuthen, Germany
\and Istituto Nazionale Fisica Nucleare (INFN), 00044 Frascati (Roma) Italy
\and Max-Planck-Institut f\"ur Physik, D-80805 M\"unchen, Germany
\and Institut de F\'isica d'Altes Energies (IFAE), The Barcelona Institute of Science and Technology (BIST), E-08193 Bellaterra (Barcelona), Spain
\and Universit\`a di Padova and INFN, I-35131 Padova, Italy
\and Universit\`a di Pisa, and INFN Pisa, I-56126 Pisa, Italy
\and ICRANet-Armenia at NAS RA, 0019 Yerevan, Armenia
\and Centro de Investigaciones Energ\'eticas, Medioambientales y Tecnol\'ogicas, E-28040 Madrid, Spain
\and Universit\"at W\"urzburg, D-97074 W\"urzburg, Germany
\and Finnish MAGIC Consortium: Finnish Centre of Astronomy with ESO (FINCA), University of Turku, FI-20014 Turku, Finland; Astronomy Research Unit, University of Oulu, FI-90014 Oulu, Finland
\and Departament de F\'isica, and CERES-IEEC, Universitat Aut\`onoma de Barcelona, E-08193 Bellaterra, Spain
\and Japanese MAGIC Consortium: ICRR, The University of Tokyo, 277-8582 Chiba, Japan; Department of Physics, Kyoto University, 606-8502 Kyoto, Japan; Tokai University, 259-1292 Kanagawa, Japan; RIKEN, 351-0198 Saitama, Japan
% ...
}

\date{Received XX XX 2019 / accepted XX XX 2020}


 
  \abstract
   {}  
   {In the presence of a sufficient amount of target material, $\gamma$ rays can be used as a tracer in the search of sources of Galactic cosmic rays (CRs). Here we present deep observations of the Galactic Centre (GC) region with the MAGIC telescopes, which we use for inferring the underlying CR distribution, and for studying the alleged PeV proton accelerator (PeVatron) at the centre of our Galaxy.}
   {We use data from ${\approx}100$~hr observations of the GC region conducted with the MAGIC telescopes over 5 years (from 2012 to 2017). Those were collected at high zenith angles (58-70~deg), leading to a larger energy threshold, but also an increased effective collection area compared to low zenith observations. Using recently developed software tools, we derive the instrument response and background models required for extracting the diffuse emission in the region. We use existing measurements of the gas distribution in the GC region to derive the underlying distribution of CRs. We further discuss the associated biases and limitations of such an approach.} 
   {We obtain a significant detection for all four model components used to fit our data (Sgr~A*, ``Arc'', G0.9+0.1, and an extended component for the Galactic Ridge). We observe no significant difference between the $\gamma$-ray spectra of the immediate GC surrounding, which we model as a point source (Sgr~A*), and the Galactic Ridge. The latter can be described as a power-law with index 2 and an exponential cut-off at around 20~TeV with the significance of the cut-off being only 2~$\sigma$. The derived cosmic-ray profile hints to a peak at the GC position and with a measured profile index of $1.2 \pm 0.3$ is consistent with the $1/r$ radial distance scaling law, that supports the hypothesis of a CR accelerator at the GC. We argue that the measurements of this profile are presently limited by our knowledge of the gas distribution in the GC vicinity.}
   {}

  
  \titlerunning{MAGIC observations of the diffuse emission in the Galactic Centre region}
  \maketitle


\section{Introduction}

The Galactic Centre (thereafter GC) is one of the most extraordinary regions in the very high energy (VHE, >100~GeV) $\gamma$-ray sky, containing a rich variety of sources, capable of accelerating charged particles and thereby producing VHE $\gamma$-ray emission~\citep{vanEldik201545, aharonian2006hess, hess_Galactic_ridge}. If those $\gamma$ rays are produced in hadronic interactions, their sources may also contribute to the overall cosmic-ray ``sea'', filling the region with energised charged particles. This sea of cosmic rays (CRs) is believed to be responsible for the diffuse Galactic plane emission detected by EGRET~\citep{hunter_egret_1997} and later studied by \textit{Fermi}/LAT~\citep{abdo_spectrum_2010}, as well as the central $\gamma$-ray ridge detected with all major Imaging Atmospheric Cherenkov Telescopes~\citep[IACTs,][]{hess_Galactic_ridge,2016VERITASGC_Smith,MAGIC_GC_2017}. The detected $\gamma$ rays originate primarily from deep inelastic nuclear interactions of the high-energetic particles with the surrounding gas, which can also be traced via its emission in the radio band (CS emission, \citealt{Tsuboi_GC_CSmaps}; CO emission, \citealt{oka_largescale_1998}). Thus deep $\gamma$-ray observations in the TeV energy range can be used to infer the underlying CR density, provided that the gas distribution is reliably measured.

This idea has been applied using observations of the central ${\approx}\pm 1^\circ$ region (in latitude) of the Galaxy (corresponding to $\approx$300~pc at a distance of 8.5~kpc) with the High Energy Stereoscopic System (H.E.S.S.), suggesting an inhomogeneous distribution of CRs around the GC~\citep{HESS_VHE_GC_ridge, hess_collaboration_acceleration_2016, HESS_GalDiffuse_2018}. Furthermore, those data indicate that the central super-massive black hole of our Galaxy may itself accelerate particles to PeV ($10^{15}$~eV) energies.

Observing the GC region with the MAGIC telescopes is only possible at large zenith distances ($>58^\circ$), implying an increased optical thickness of the atmosphere and larger distance to the shower maximum, which leads to a stronger dilution of the Cherenkov light. On the other hand, for geometric reasons, these data benefit from an increased collection area~\citep[${\approx}1~\mathrm{km^2}$,][]{MAGIC_GC_2017} and thus a boost of sensitivity at energies above several TeVs.

\citet{MAGIC_GC_2017} have already presented the first part of the MAGIC GC observation campaign in 2012-2015, consisting of ${\approx}70$~hours of exposure. Apart from the detection of multi-TeV emission from the Galactic plane, those data suggested the presence of a new VHE source close to the so-called radio ``Arc'' in the GC vicinity, detected also by other telescopes~\citep{2016VERITASGC_Smith,HESS_GalDiffuse_2018}. We continued observing the GC during 2015 - 2017, adding up to a total exposure time of ${\approx}100$~hours. In addition, we employ the recently developed 2D likelihood analysis package SkyPrism~\citep{skyprism}, which provides a more sensitive analysis of extended sources for MAGIC.

This paper is organised as follows. In Sect.~\ref{sect::magic_observations_general} we describe the MAGIC observations, the data selection strategy and the data analysis techniques. In Sect.~\ref{sect::data_analysis_general} we provide a detailed description of the various analysis steps and report on their results. In Sect.~\ref{sect::biases_general} we estimate the biases, stemming from our background modelling and analysis approaches. Finally, Sects.~\ref{sect::discussion} and~\ref{sect::conclusions} provide a general discussion and a summary of the obtained results.

\section{MAGIC observations of the Galactic Centre region}
\label{sect::magic_observations_general}


The MAGIC (Major Atmospheric Gamma Imaging Cherenkov) telescopes are two 17~m diameter IACTs, located at a distance of 85~m from each other at an altitude of 2200~m a.s.l. at the Roque de los Muchachos Observatory on the Canary Island of La Palma, Spain (28$^\circ$N, 18$^\circ$W).

The telescopes record flashes of Cherenkov light produced by Extensive Air Showers (EAS) initiated in the upper atmosphere by $\gamma$-ray photons within a field of view of $1.5^\circ$ (in radius). Both telescopes are operated together in a coincidence trigger stereoscopic mode, in which only events simultaneously triggering both telescopes are recorded and analysed~\citep{Magic_performanceII}. For low zenith distance observations and for \mbox{$E>220$~GeV}, the integral sensitivity of MAGIC is $(0.66\pm0.03)\%$ in units of the Crab Nebula flux (C.U.) for 50 hours of observation~\citep{Magic_performanceII}. At larger zenith angles (above $60^\circ$), at which the GC is observable at La Palma, the MAGIC collection area reaches 1~km$^2$ at the energies above 10~TeV~\citep{MAGIC_GC_2017}.


The MAGIC observations of the GC region, used for this study, were carried out between April 2012 and May 2017 (typically from March to July). The observation time after quality selection cuts is ${\approx}100$~hr. The data were taken in the so-called ``Wobble'' mode~\citep{1994Fomin_wobble}, centred on Sgr~A* with a pointing offset of $0.4^\circ$.


The acquired data have been processed with the standard MAGIC Analysis and Reconstruction Software~\citep[MARS,][]{zanin2013mars}. These steps include selection of good data quality periods, the low-level processing of the camera images, the calculation of image parameters and the reconstruction of the energy and incoming direction of the primary $\gamma$ rays. The last step is performed using the machine learning technique Random Forests \citep{Magic_RF}, where Monte Carlo events and real background data are used for training, and which is also used for background suppression through event classification.

At the quality selection step, we removed events corresponding to known temporary hardware issues or recorded during bad weather periods. As the full data sample contains some nights with weak moonlight, we also exclude time periods with sky brightness $\gtrsim 2.5$ times the typical dark night sky brightness. Data recorded with higher sky brightness lead to biases when analysed with the same pipeline that is used for data that has been recorded during dark nights, but can in principle still be used when accepting a higher energy threshold when using special analysis settings \citep[see][]{ahnen_performance_2017}. In order to select only good quality data we applied cuts on several instrument and weather related parameters. Those include the mean photomultiplier currents, the event trigger rate, the number of stars detected by the MAGIC star-guider cameras and the cloud-/aerosol- induced absorption in the telescope's field of view. The latter is derived from measurements with an infra-red pyrometer and a LIDAR system~\citep{gaug_atmospheric_2014,fruck_novel_2014}.


\section{Data analysis}
\label{sect::data_analysis_general}

The high-level analysis of the acquired data has been performed with the set of utilities available in the MAGIC SkyPrism package~\citep{skyprism}. SkyPrism provides a set of routines, aimed at 2D spatial likelihood analysis of MAGIC data, which allows us to self-consistently account for sources of complex morphology within the field of view. Furthermore, it also incorporates algorithms for computing the instrument point spread function, exposure and background maps, which we intensively use throughout the analysis presented here.


\subsection{Diffuse background model construction}

The wobble observational scheme~\citep{1994Fomin_wobble}, used here, 
provides a simultaneous background estimate from comparison of several adjunct (albeit exposed to slightly different sky regions) positions. Still the applicability of this approach is limited by the distance between the wobble pointings and the target, which in our case was $d_w = 0.4^\circ$. In practice, a source with an extension comparable to or larger than twice the offset distance ($0.8^\circ$), in the same direction, will still (partially) contribute to the background measured in the same camera region. The diffuse emission of the Galactic plane clearly exceeds this limit, so a special treatment is necessary.

Such contamination of the background map with excess $\gamma$-ray signal can be avoided if contributions from locations close to known $\gamma$-ray sources are excluded during map construction~\citep{skyprism}. We thus mask out camera regions that correspond to the Galactic plane and use the rest of the camera (free from known/expected sources) to estimate the corresponding background. This approach, implemented in the \textit{SkyPrism} package, allows us to significantly reduce the background bias.

Due to the remaining limitations of our background modeling technique and the used wobble scheme, we can not completely remove the possible bias in the background estimation from a source as extended as the diffuse Galactic plane emission. Our estimates described in detail in Sect.~\ref{sect::background_bias} and illustrated in Fig.~\ref{fig::bkg_bias_map} demonstrate that in the $l=[-1.5^\circ;1.5^\circ]$ longitude range along the plane the remaining bias is within 10-20\% of the assumed local source luminosity throughout the Galactic plane and does not exceed 30-50\% (equivalent to ${\approx}3\%$ of background) in the outskirts. At the same time the flux bias is rather constant across the plane and its variations, averaged in the latitude range $b=[-0.2^\circ;0.2^\circ]$ do not exceed ${\approx}1\%$ of the background, which corresponds to 10-20\% of the estimated source flux for a highly extended source.


\subsection{MAGIC view of the Galactic Centre region}

The large ($>58^\circ$) zenith angle GC observations imply an increased energy threshold of ${\approx}1$~TeV, though an analysis is also possible at even lower energies, given that the studied source is bright enough~\citep{MAGIC_GC_2017}. Thus we have performed the spectral analysis in the 400~GeV -- 50~TeV energy range; the morphology study of the diffuse emission in the GC region was done above $\gamma$-ray energies of 1~TeV.

\begin{figure}
  \centering
  \includegraphics[width=\columnwidth]{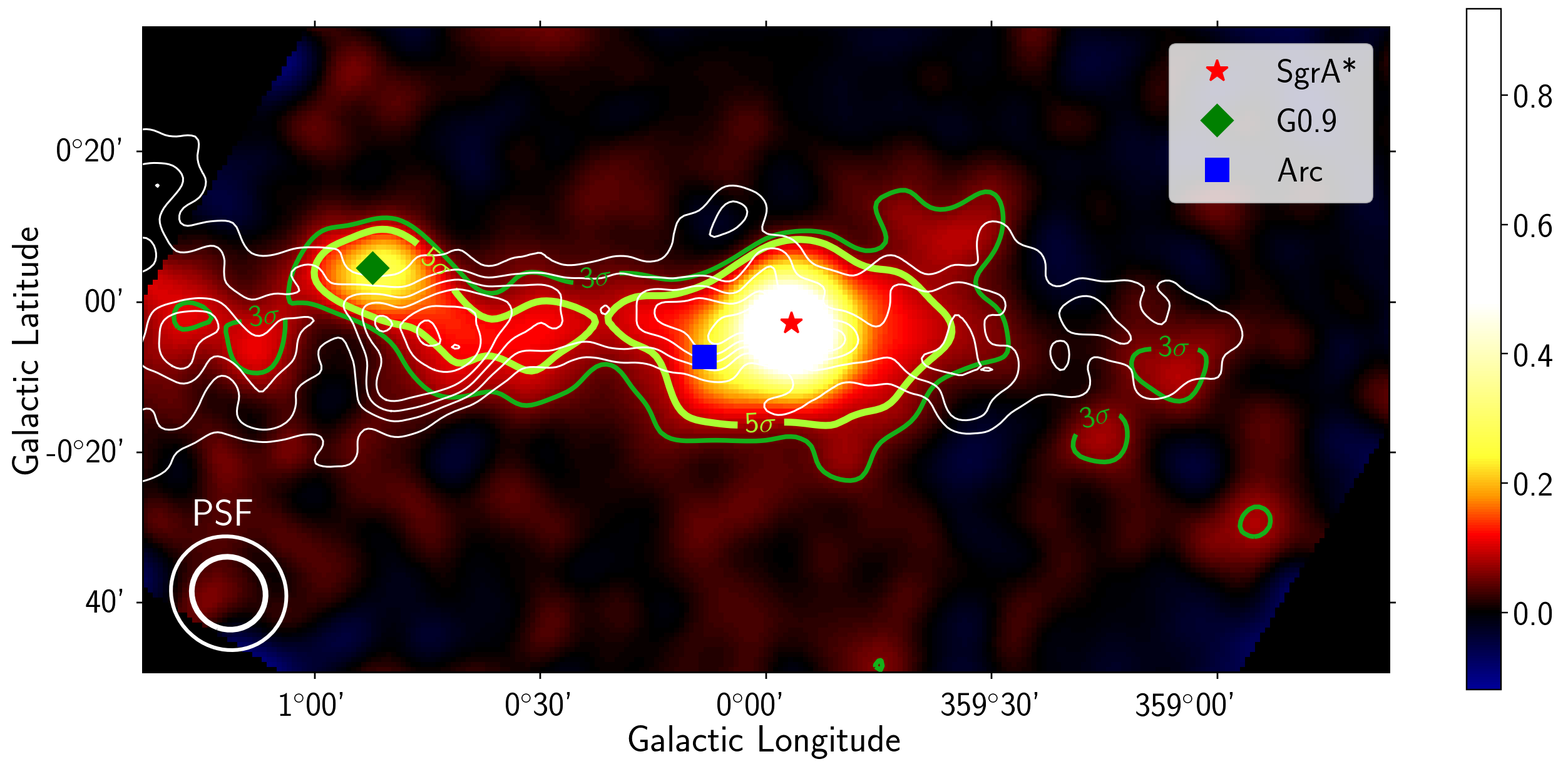}
  \caption{Sky map (excess in units of background) of the GC region in Galactic coordinates at energies above 1~TeV, smeared with a kernel resembling the MAGIC PSF. The pre-trial statistical significance of regions with excess counts is highlighted by green contours at the levels $5\sigma$ and $3\sigma$. The (smeared) MAGIC PSF is indicated by $39\%$ and $68\%$ containment contours. The white contours show radio line emission from CS molecules, tracing dense gas~\citep{Tsuboi_GC_CSmaps}.}
  \label{fig::RelFluxMap_GC_1000-20000}
\end{figure}

The sky map (above $E=1$~TeV) of the GC vicinity, produced with the described diffuse background estimation scheme, is shown in Fig.~\ref{fig::RelFluxMap_GC_1000-20000}. The Galactic plane is visible over $2^\circ$ across the image. The significance of this detection, computed using the CS radio emission profile \citep{Tsuboi_GC_CSmaps} as an approximation (see Sect.~\ref{sect::gal_plane_scan} and~\ref{sect::spectra} for details of the method), results in a ${\approx}17$ standard deviations ($\sigma$) incompatibility with the null hypothesis of background and point sources only. Other sources visible in the image are Sgr~A*, G0.9+0.1, and the so-called ``Arc'', detected with significances of ${\approx}48\,\sigma$, ${\approx}11\,\sigma$ and ${\approx}6.4\,\sigma$ respectively, propagating also uncertainties of the background and exposure model.


\subsection{Galactic plane brightness scan}
\label{sect::gal_plane_scan}

The CR distribution profile in the GC surroundings should roughly correspond to the brightness distribution of the detected $\gamma$-ray emission. Previous measurements have already shown evidence that the $\gtrsim 100$~GeV brightness of the Galactic plane is peaking towards the Sgr~A*, indicating a concentration of cosmic-rays~\citep{hess_collaboration_acceleration_2016}. The presence of an extended central component, reported by  the ~\citet{HESS_GalDiffuse_2018}, also speaks in favour of this assumption.

The cosmic-ray distribution around the GC can be inferred by solving the integral equation
\begin{equation}
  S(x,y) = A \int \rho_{gas}(x,y,z) \, \rho_{CR}(x,y,z) \, dz\,,
  \label{eq::image_brightness}
\end{equation}
where $S(x,y)$ is the image plane $\gamma$-ray brightness distribution, $\rho_{gas}$ and $\rho_{CR}$ are the number densities of the gas and CRs respectively and $A$ is a factor, which takes into account the proton-proton interaction cross-section, distance to observer and additional constants. A proper solution of this equation requires knowledge of the full 3D gas density distribution $\rho_{gas}(x,y,z)$, which is challenging to obtain. Indeed, while in the image plane ($x,y$) the resolution of radio surveys reaches ${\approx}0.01^\circ$~\citep[for instance the CS $J=1-0$ emission radio survey of][]{Tsuboi_GC_CSmaps} equivalent to 1-2~pc scale, the line-of-sight distance $z$ can hardly be obtained with an accuracy better than several tens of parsecs. For this reason we solve an approximate expression of Eq.~\ref{eq::image_brightness}:
\begin{equation}
  S(x,y) \approx A \int \rho_{gas} \, dz \times \int \rho_{CR} \, dz = P_{gas}(x,y) \,  P_{CR}(x,y)\,,
  \label{eq::image_brightness_approx}
\end{equation}
which splits the problem into the projected gas (directly inferred from the radio data) and cosmic-ray distributions $P_{gas}(x,y)$ and $P_{CR}(x,y)$ respectively. To avoid degeneracy, we just consider radially symmetric cosmic-ray profiles $\rho_{CR} \propto r^{-\alpha}$ (with $r$ being the distance from the GC) and their projections onto the image plane. These simplifications result in a certain bias of our measurement, which we quantify in Sect.~\ref{sect::model_bias}.

\begin{figure}
  \centering
  \includegraphics[width=\columnwidth]{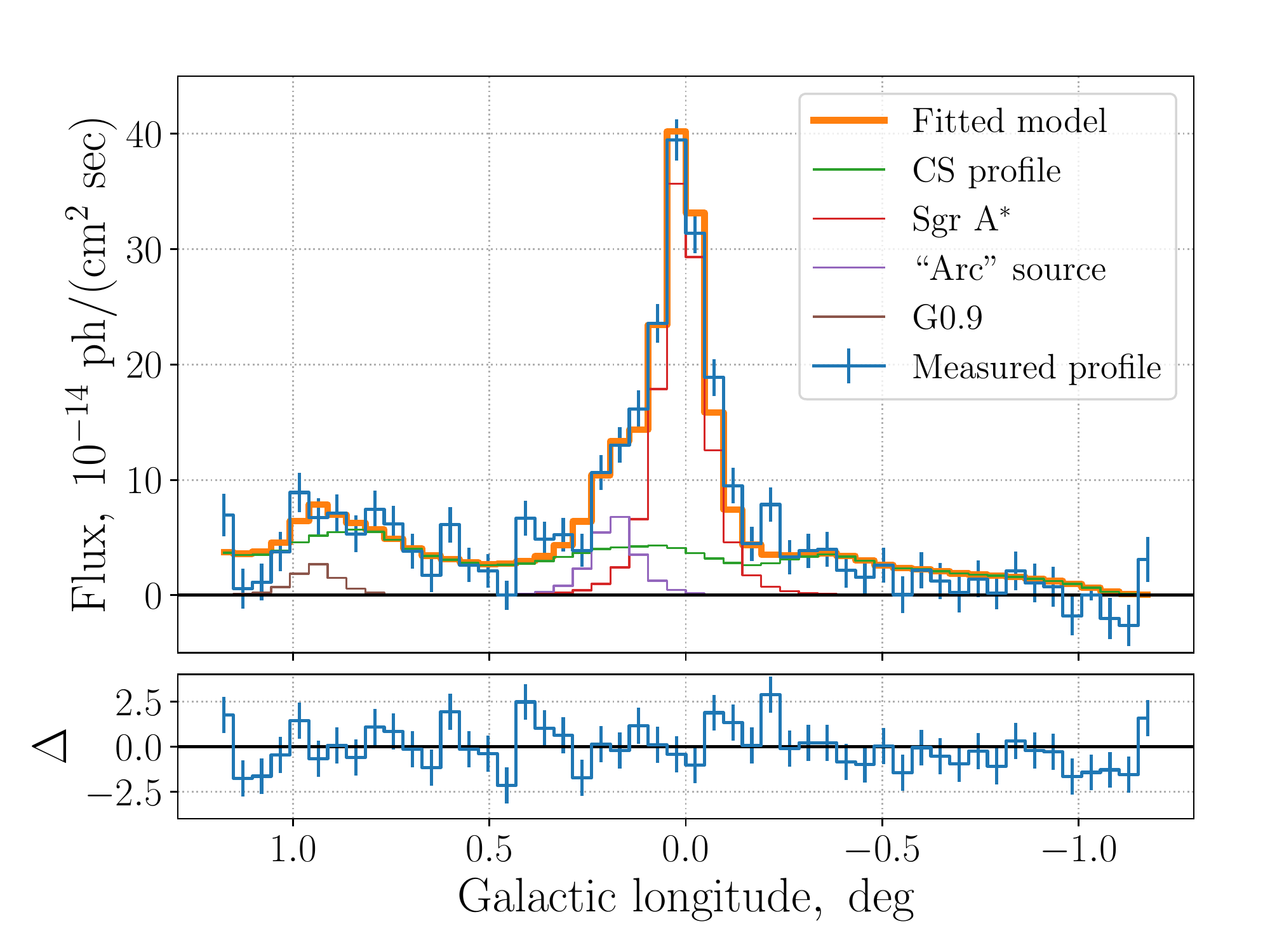}
  \caption{
    \textit{Top}: Brightness scan of the $b=[-0.2^\circ;0.2^\circ]$ stripe of the Galactic plane in the energy range above 1.2~TeV. Blue entries denote the MAGIC measurements, whereas the orange line is the best-fit model to them, composed of the CS profile, Sgr~A* point source and the extended ``Arc''.
    \textit{Bottom}: Residuals of the fit in the units of measurement uncertainties.
  }
  \label{fig::GalPlane_scan_above_1200GeV}
\end{figure}

First we test whether a homogeneous cosmic-ray distribution $\rho_{CR} = const$ is consistent with the MAGIC data. For this we have produced an excess event profile of the $b=[-0.2^\circ; 0.2^\circ]$ stripe, centred at the GC position, and the corresponding MAGIC exposure profile using the features of SkyPrism. The resulting, exposure-normalized, Galactic plane profile above 1.2~TeV is shown in Fig.~\ref{fig::GalPlane_scan_above_1200GeV}. Fitted with a simple model, containing the Sgr~A* and G0.9 point sources, the ``Arc'' source,
and the  diffuse emission model $S(x,y)$ (computed from the CS emission maps with $\rho_{CR} = const$), it results in $\chi^2/d.o.f. \approx 69/46$ degrees of freedom, equivalent to ${\approx}2.4\sigma$ disagreement of the data with the model.

In order to investigate if the MAGIC data are in a better agreement with a $\rho_{CR} \neq const$ type cosmic-ray profiles we use a grid search for the optimal value of $\alpha$ using the profile shown in Fig.~\ref{fig::GalPlane_scan_above_1200GeV} and estimated the cosmic-ray density $P_{CR}(d) = S(d)/P_{gas}(d)$ as function of the projected off-centre distance $d$ using a full maximum likelihood fit of the measured $\gamma$-ray  brightness around the GC.

\begin{figure}
  \centering
  \includegraphics[width=\columnwidth]{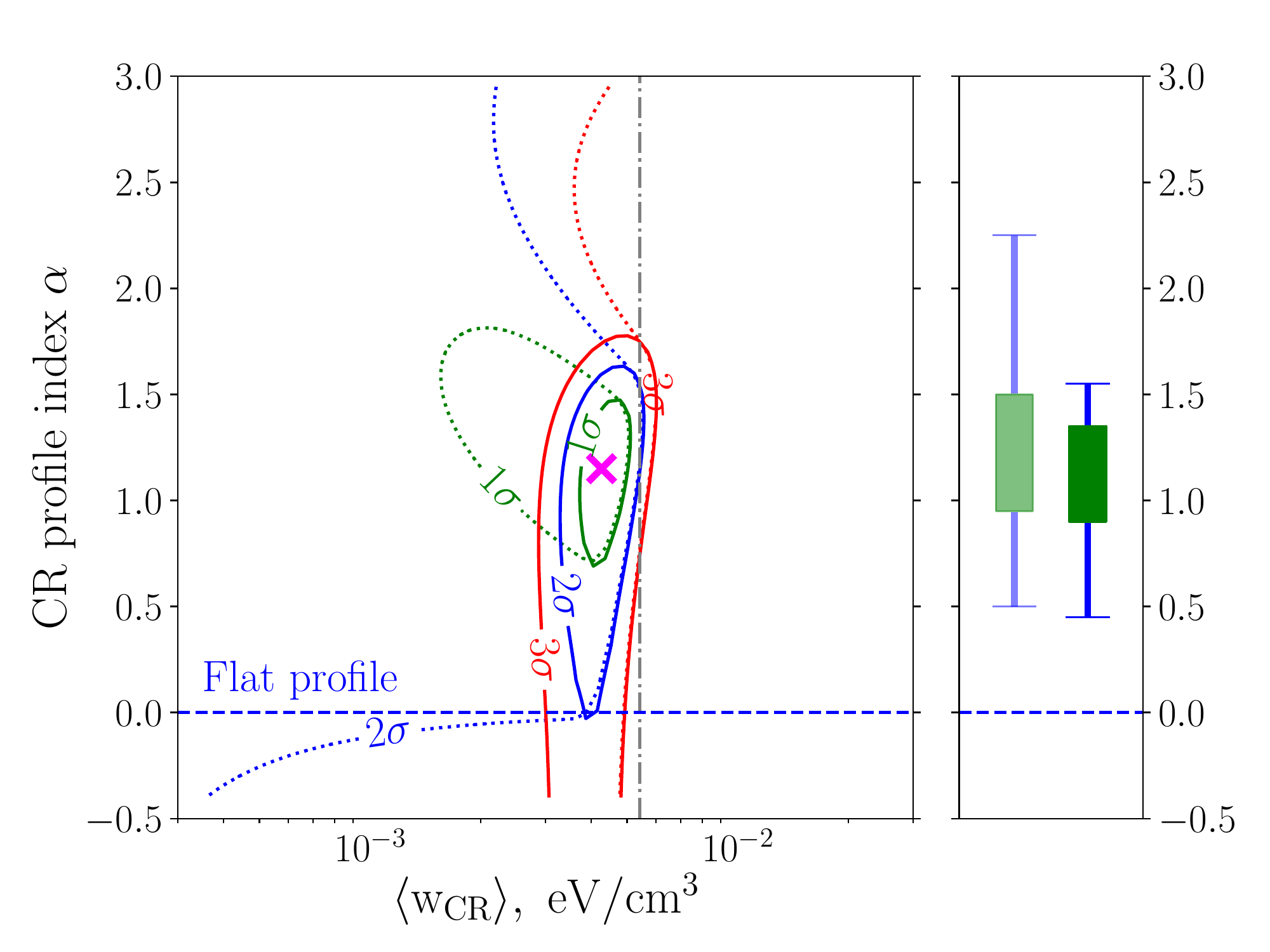}
  \caption{ 
    \textit{Left}: Likelihood scan of the cosmic-ray density profile parameter space for $E_\mathrm{CR} \gtrsim 10$~TeV, for a centrally-peaked profile of the form $\rho_{CR} \propto 1/r^\alpha$, based on the MAGIC measurements. The scan is performed for two assumptions: (1) only the cosmic-ray population with the power-law distribution exists (solid lines) and (2) the power-law population exist on top of a uniform cosmic-ray density distribution (dashed lines). Magenta crosses mark the best fit values for both assumptions, which nearly exactly overlap for our data. Gray vertical dash-dotted line represents the mean cosmic ray density from~\citep{hess_collaboration_acceleration_2016}. The mean cosmic-ray energy density $\left< w_{CR} \right>$ estimated here corresponds to the Galactic longitude range $[-1^\circ;1^\circ]$.
    \textit{Right}: Marginalization of the scan over the cosmic-ray profile index only. Green boxes and blue error bars mark the 1 and $2 \sigma$ confidence ranges correspondingly for assumptions 1~(dark colours) and 2~(light colours).
   }
  \label{fig::CR_power_scan}
\end{figure}

The results of the likelihood-profile scan are shown in Fig.~\ref{fig::CR_power_scan}, where the density $\rho_{CR}$ is converted to the cosmic-ray energy density $w_{CR}$ using the same procedure as in~\citet{hess_collaboration_acceleration_2016} (formula (2) of the methods section):
\begin{equation}
  w_{CR}(\geq 10 E_\gamma) \approx 1.8 \cdot 10^{-2} 
  \left(\frac{\eta_N}{1.5}\right)^{-1} 
  \left(\frac{L_\gamma(\geq E_\gamma)}{10^{34}~\mathrm{erg/s}}\right) 
  \left(\frac{M}{10^6 M_{\astrosun}}\right)^{-1}\,,
\end{equation}
where $w_{CR}$ is in eV/cm\textsuperscript{3}, $\eta_N \approx 1.5$ accounts for nuclei heavier than hydrogen, in both CRs and in the target gas. For estimation of the H$_2$ target mass $M$ based on the CS radio maps we used the procedure described by the authors in section 4.2 of \citet{Tsuboi_GC_CSmaps}:
\begin{align}
    \begin{split}
        M(H_2)[M_{\astrosun}] =& \ \ 7.5 \cdot 10^{11} T_{ex} [\mathrm{K}] \ \cdot \\
        \cdot& \frac{ \int T_{MB}\ \mathrm{d}v [\mathrm{K\ km\ s^{-1}}] 
        \cdot A[\mathrm{cm}^2] 
        \cdot \mu(\mathrm{H}_2)[M_{\astrosun}] } 
        { X(\mathrm{CS}) }
    \end{split}
\end{align}
where the excitation temperature $T_{ex}$ of CS is 30~K, $\int T_{MB}\ \mathrm{d}v$ is the measured velocity integrated antenna temperature, $A$ is the area of the GC region, $\mu(\mathrm{H}_2)$ is the mass of the hydrogen molecule and $X(CS) = 10^{-8}$ is the relative abundance of CS in H$_2$ clouds.

In the scan we have tested two different scenarios: (1) the $\rho_{CR} \propto r^{-\alpha}$ profile dominates the cosmic-ray density in the region and (2) the peaked cosmic-ray profile is found on top of an underlying homogeneous density, so that $\rho_{CR} \propto r^{-\alpha} + const$. To account for the background and telescope exposure uncertainties, we have generated 50 random exposure and background maps representing the uncertainty range of our reconstruction method and repeated the scan for each of them. We then averaged the resulting likelihood values in each bin of the  scan phase space, which is equivalent to marginalization over these random map representations. The same technique of propagating the uncertainties of the background and exposure models has also been applied throughout the spectral analysis (Sec. \ref{sect::spectra}). The resulting averaged values were then used to compute the confidence contours, shown in Fig.~\ref{fig::CR_power_scan}.
In addition to $\alpha-w_{CR}$ combined confidence contours, the right panel of the same figure shows the marginalised uncertainties for the power-law index $\alpha$ for both scenarios (1) and (2), with darker and lighter colours respectively.

Scenario (1) favours a cosmic-ray density profile with $\alpha = 1.2_{-0.3}^{+0.2}$ ($1\sigma$ uncertainty). A similar profile with $\alpha = 1.2 \pm 0.3$ is found for scenario (2), where a homogeneous contribution to the cosmic-ray profile is allowed.

\begin{figure}
  \centering
  \includegraphics[width=\columnwidth]{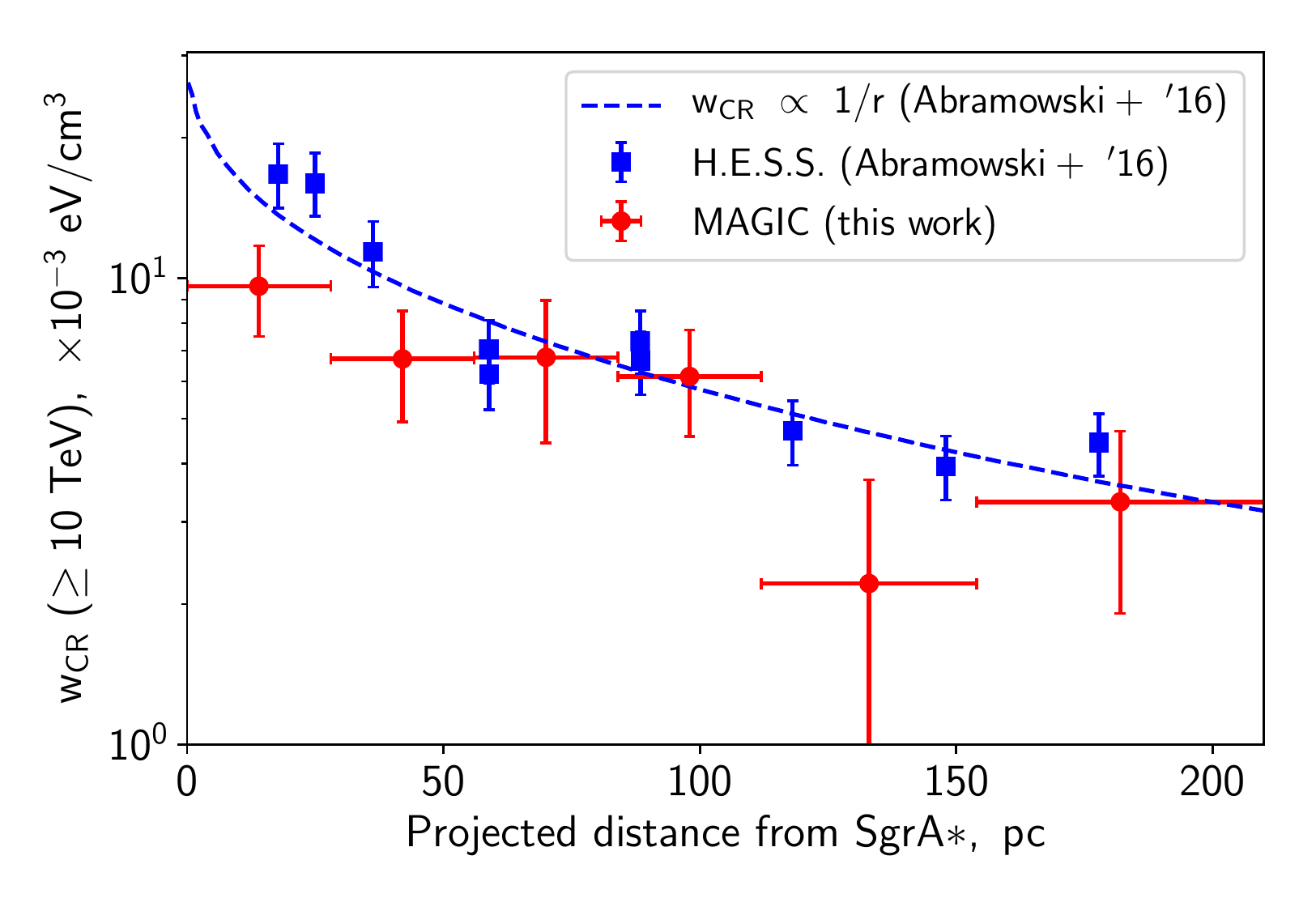}
  \caption{The projected cosmic-ray energy density, as obtained from the full likelihood fit to the MAGIC sky map above 1~TeV. The projected distance is counted from the GC position. Measurements of~\citet{hess_collaboration_acceleration_2016} are shown in blue for comparison.}
  \label{fig::CR_power_profile}
\end{figure}

As it is illustrated in Fig.~\ref{fig::CR_power_profile}, also the full likelihood fit to the obtained sky map is not consistent with the $\rho_{CR} = const$ assumption. To perform this fit, we have first split the CS emission map of~\citet{Tsuboi_GC_CSmaps} (integrated over the radial velocity) into a sequence of concentric rings, with their own normalisation factors. Since the CS emission is highly peaked toward the Galactic plane, this is effectively equivalent to a longitudinal split of the Plane inside the $b=[-0.1^\circ, 0.1^\circ]$ stripe. When computing the normalisations of the rings, we've also included the Sgr~A* point source, as well as G0.9+0.1 and the ``Arc'' source to the fitted model. The resulting cosmic-ray density profile, shown in Fig.~\ref{fig::CR_power_profile} was finally fit with the $\rho_{CR} = const$ model, yielding $\chi^2\approx 22$ over 5 degrees of freedom. This corresponds to ${\approx}3.5\sigma$ data to model disagreement, indicating a peaked profile.

In summary, MAGIC data above 1~TeV indicate a radial cosmic-ray profile with power index $\alpha \approx 0.9-1.4$, somewhat different but still compatible to earlier findings by~\citet{hess_collaboration_acceleration_2016}.


\subsection{Spectral analysis of the detected sources}
\label{sect::spectra}

To compute energy spectra of the detected sources in the MAGIC field of view -- including the diffuse Galactic plane emission -- we have used the the SkyPrism package. The spatial model used for the fit includes three point sources (Sgr~A*, G0.9+0.1 and the Arc source at RA=17:46:00, Dec=-28:53:00) and the velocity integrated CS map, re-scaled with the $\rho_{CR} \propto r^{-1.2}$ best-fit cosmic-ray profile. The fit has been performed in the energy range from 400~GeV to 50~TeV with two methods -- energy bin-wise (7 logarithmic energy bins) and assuming a certain spectral shape model for each of the sources. In the latter case we applied a forward folding procedure considering the energy migration matrix during the fit. All spectra were generated from the general form
\begin{equation}
  dN/dE = N \left( \frac{E}{E_0} \right)^{ (\Gamma + \beta \log{(E/E_0)}) } \exp{(-E/E_{cut})}\,,
\end{equation}
which can result in a power-law, log-parabola and power-law with cut-off spectral shape depending on the choice of $\beta$ and $E_{cut}$. The normalisation energy for all the sources was set to $E_0 = 2$~TeV, keeping the correlation between the spectral parameters minimal.

\begin{figure}
  \centering
  \includegraphics[width=1.0\columnwidth]{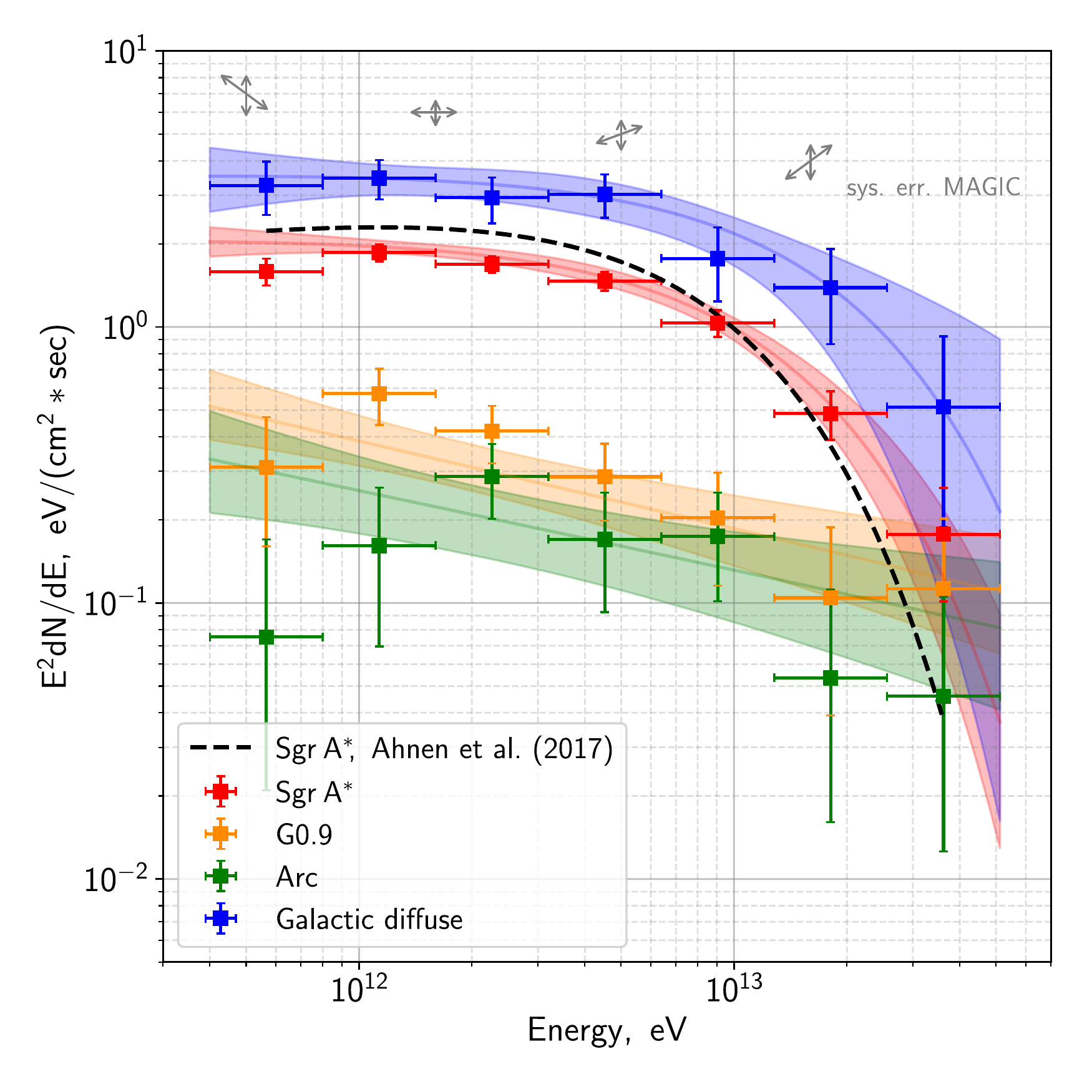}
  \caption{MAGIC SEDs of the different components in our model, data points and forward folding fit results (colored bands). For the components corresponding to Sgr~A*, the ``Arc'' and the CR/MC component a power-law shape with exponential cut-off has been used while G0.9+0.1 can be described with a simple power-law. The error bars and bands were computed from the MCMC samples and correspond to 68\% confidence range. No spectral unfolding has been applied to the data points, but the effect of spillovers due to energy migration has been corrected for, based on the spectral shapes. Gray arrows indicate the size (length) and direction (orientation) of the SED shifts due to the systematical uncertainties in the analysis (see Sect.~\ref{sect::spectra} for details).}
  \label{fig::SED_all}
\end{figure}

\begin{figure}
  \centering
  \includegraphics[width=\columnwidth]{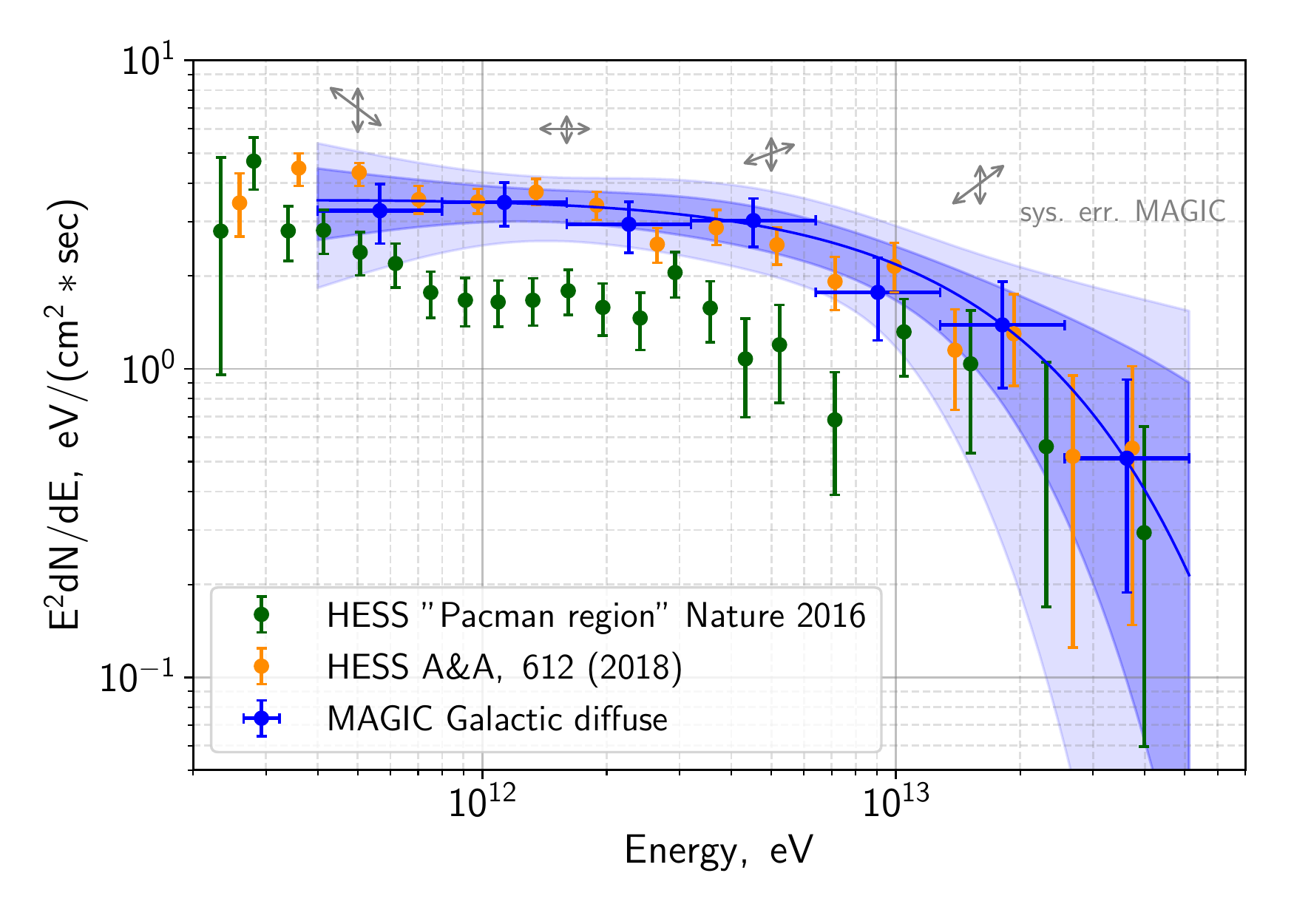}
  \caption{The spectrum of the diffuse Galactic emission, derived from the MAGIC data. Dark and light blue regions mark the 68\% and 95\% confidence ranges for the assumed power-law with exponential cut-off model. The diffuse spectrum from~\citet{HESS_GalDiffuse_2018}, extracted from a similar region, is shown in orange, while the SED obtained from a cut annulus with 0.45 deg outer radius from~\citet{hess_collaboration_acceleration_2016} is shown in green. Gray arrows indicate the possible shifts due to the systematical uncertainties in the analysis, similar to Fig.~\ref{fig::SED_all}.}
  \label{fig::diffuse_spectrum}
\end{figure}

To obtain the best fit parameters of the assumed spectral models we perform a maximal Poissonian likelihood fit to the energy-binned MAGIC sky maps. To ensure a good accuracy of the estimated uncertainties, we have additionally used the Markov Chain Monte Carlo (MCMC) sampler \textit{emcee} on the parameter space~\citep{emcee}. The uncertainties of the exposure and background models that were derived from Monte Carlo simulations and data regions more than $0.3\deg$ off the Galactic plane, were propagated to the final results by processing 60 random representations through the MCMC sampler and merging the samples. The best-fit values for the detected sources, obtained through this fit, are given in Tab.~\ref{tab::spec_results}, along with the corresponding errors and detection significances. The obtained spectra (data points and fit results) are shown in Fig.~\ref{fig::SED_all}. The data points are not the result of spectral unfolding, but spillover corrections based on the energy migration matrix and the fitted spectral shape were applied. The obtained MAGIC spectrum is consistent with the earlier estimate of the Galactic ridge Spectral Energy Density (SED)~\citep{HESS_GalDiffuse_2018}, as displayed in Fig.~\ref{fig::diffuse_spectrum}. A likelihood ratio test comparing the model for the diffuse component with cut-off to a pure power-law results in the ${\approx}2\,\sigma$ preference for the cut-off for the MAGIC data set. 

The SED shown earlier by~\citet{hess_collaboration_acceleration_2016} that led to the speculation about a possible PeVatron at the GC, shows a lower average flux and somewhat different spectral shape compared to the other two SEDs in Fig.~\ref{fig::diffuse_spectrum}. This difference could be explained by the fact that also the regions in which the fluxes were measured are different. While~\citet{HESS_GalDiffuse_2018} and this work try to include the whole $< 1$~deg from the GC part of the Galactic ridge, avoiding point sorces, \citet{hess_collaboration_acceleration_2016} used a donut shaped region for extracting their flux, with a cut-out at the position of the Arc source and inner and outer radii of 0.15~deg and 0.45 deg respectively.

We estimate the systematic uncertainties, arising from uncertainties on the energy and flux normalization scales, following the procedure discussed in~\citet{MAGIC_GC_2017} \citep[based on a detailed study by][]{Magic_performanceII}. The resulting estimates are indicated by gray arrows in Fig.~\ref{fig::SED_all} and Fig.~\ref{fig::diffuse_spectrum}, where the vertical arrows indicate the effect of the flux normalization errors at different energies and the horizontal or inclined arrows indicate the effect of the energy scale uncertainty. 

\begin{table*}
   \center
   \small
   \caption{
     Table with the spectral fit results of the sources, detected in the MAGIC field of view. The spectrum type acronyms stand for: PL -- power-law and PLC -- cut-off power-law. Normalisation factor $N$ is given in units of $10^{-25}~\mathrm{[ph/(cm^2~s~eV)]}$; normalisation energy is set to $E_0=2~\mathrm{TeV}$ for all the sources. The curvature parameter $\beta=0$ in all cases. For the sources fitted with the power-law model the values of $E_{cut}$ are not given. All uncertainties correspond to a $68\%$ confidence interval.
   }
   \linespread{1.75}\selectfont  
   \begin{tabular}{cccccccc}
      \hline
      \hline
      Name     & Spec. type      & N, $10^{-25}$ eV$^{-1}$ cm$^{-2}$ s$^{-1}$ & $\Gamma$   & $E_{cut}$, TeV            & Detection significance \\
      \hline
      Sgr~A*   & PLC           & $5.39_{-0.46}^{+0.56}\,$stat.$_{-1.19}^{+1.61}\,$sys. & $-1.98_{-0.10}^{+0.11}\,$stat.$_{-0.17}^{+0.18}\,$sys. & $12.4_{-3.2}^{+5.5}\,$stat.$_{-0.2}^{+3.3}\,$sys.   & $\approx48\ \sigma$         \\
      G0.9+0.1 & PL            & $0.93_{-0.17}^{+0.20}\,$stat.$_{-0.12}^{+0.26}\,$sys. & $-2.32_{-0.15}^{+0.13}\,$stat.$_{-0.12}^{+0.20}\,$sys. & --                                                  & $11.1\ \sigma$           \\
      Arc      & PL            & $0.52_{-0.15}^{+0.15}\,$stat.$_{-0.09}^{+0.16}\,$sys. & $-2.29_{-0.19}^{+0.17}\,$stat.$_{-0.13}^{+0.23}\,$sys. & --                                                  &  $6.4\ \sigma$           \\
      Diffuse  & PLC           & $9.32_{-1.63}^{+2.39}\,$stat.$_{-1.97}^{+2.53}\,$sys. & $-1.98_{-0.21}^{+0.26}\,$stat.$_{-0.15}^{+0.16}\,$sys. & $17.5_{-9.55}^{+59.3}\,$stat.$_{-1.9}^{+4.5}\,$sys. & $17.3\ \sigma$           \\
      \hline
   \end{tabular}\\
   \label{tab::spec_results}
   
\end{table*}

\section{Estimation of the possible biases in the analysis}
\label{sect::biases_general}


\subsection{Bias from the background modelling}
\label{sect::background_bias}

In order to quantify the bias resulting from the background estimation, we used a simplified simulation of the background map based on the initial assumption on the extension and brightness of the sources in the GC region. In this simulation we assume that the true signal measured by MAGIC consists of five contributions: extended gas emission (assumed to be traced by the CS map~\citep{Tsuboi_GC_CSmaps}), point-like Sgr~A*, the ``Arc'' source~\citep{2016VERITASGC_Smith,MAGIC_GC_2017,HESS_GalDiffuse_2018}, point-like G0.9+0.1 and isotropic background. The relative normalisations of these components are taken from the previous analysis of~\citet{MAGIC_GC_2017};  we used our best fit results for a cross-check. This composite image is then used to sample the photons in the telescope camera coordinates as a function of pointing azimuth and zenith, following the MAGIC pointing during the GC observations. The resulting event list is supplied to the background estimation routine for a comparison of the reconstructed vs. assumed background.

\begin{figure*}
  \centering
  \includegraphics[width=\linewidth]{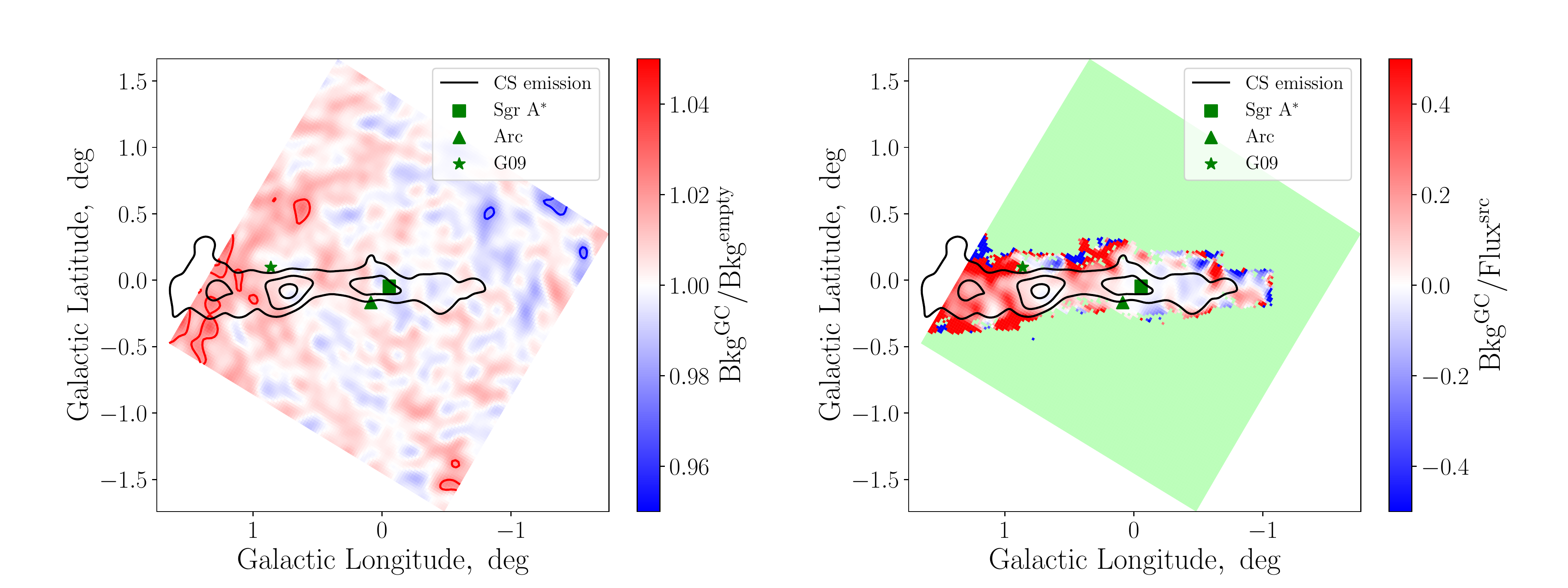}
  \caption{
    \textit{Left}: The background bias (in units of the true assumed background flux $Bkg^{empty}$) for the used background reconstruction technique, estimated with simulations. The map was smeared with the Gaussian kernel with $\sigma=0.05^\circ$; contours indicate $\pm 2\%$ differences with respect to the reference value of 1. Also plotted are the positions of the known $\gamma$-ray sources in the galactic center region as well as the CS emission contours from~\citet{Tsuboi_GC_CSmaps}, corresponding to the integral antenna temperatures of 5, 10 and 15~K, smeared with the same Gaussian kernel. 
    \textit{Right}: The same, but in units of the source flux for the assumed model. Here only the uncertainties within the region of the used CS gas map are plotted, as the rest of the modelled image contains only background. Such background-only regions are filled with light green.
  }
  \label{fig::bkg_bias_map}
\end{figure*}

Based on the results, illustrated in Fig.~\ref{fig::bkg_bias_map}, we expect the bias to stay below 2\% in units of background flux nearly everywhere in the sky-maps. This translates to a bias of the measured flux of the diffuse emission of more than 40\% in some regions along the edges of the Galactic plane, but less than 30\% for the brighter regions along the plane and at the centre. Still, using the CS map as an approximation for the $\gamma$-ray emission, the total bias on the integral flux of the Galactic plane component is estimated to be in the range 7-12\%.


\subsection{Bias and uncertainty from the assumed gas distribution model}
\label{sect::model_bias}

An accurate modelling of the $\gamma$-ray emission from GC region requires detailed knowledge of the gas distribution in three dimensions. The angular resolution of MAGIC is better than $0.1^\circ$, which translates into ${\approx}15$~pc at 8.5~kpc distance from the Earth. As a result, MAGIC can map the profile of $\gamma$-ray emission at the projected distances of tens of parsecs from the position of the central supermassive black hole (SMBH); a conversion of this profile to the cosmic-ray density distribution then naturally requires that the line-of-sight distances to the gas clouds in the region are known with similar or better accuracy.

This requirement is very difficult to fulfil in practice. At larger distances the locations of the gas clouds are generally inferred from their kinematics, assuming a certain model of gas orbital motion~\cite[e.g.][]{sofue95, nakanish03}. In the vicinity of the GC, however, this approach can no longer be applied, as the inability to put the source in front or behind the black hole image plane at the scales of tens of parsecs leads to degeneracy in the calculations. The required information -- to a certain degree -- can be reconstructed using measurements of line-of-sight absorption of the molecular cloud emission, which provides the necessary line-of-sight position estimates~\citep{Sawada_GC_face_on}. Nevertheless, the line-of-sight locations of separate clouds in the GC region can hardly be reconstructed with an accuracy better than~${\approx}50$~pc.

In the absence and therefore negligence of the line-of-sight information, Eq.~\ref{eq::image_brightness} naturally simplifies to Eq.~\ref{eq::image_brightness_approx}, which works only with the projected gas and cosmic-ray densities. Depending on the real distribution $\rho_{gas}(x,y,z)$, the transition Eq.~\ref{eq::image_brightness}\textrightarrow~Eq.~\ref{eq::image_brightness_approx} may bear an oversimplification, resulting in a biased cosmic-ray profile $\rho_{CR}(r)$. 

We do not attempt to reconstruct a fully realistic structure of the central ${\approx}200$~pc of our Galaxy, which is rather complex~\citep{ferriere07}. Still, in order to quantify the associated bias in our analysis, we reconstruct the 3D gas distribution in the GC  region based on the measurements of~\citet{Sawada_GC_face_on} and~\citet{Tsuboi_GC_CSmaps}. This reconstruction then allows us to compare the profiles obtained accounting for or neglecting the line-of-sight information.

To perform the reconstruction, we used the fact that the 2D CS gas emission images of~\citet{Tsuboi_GC_CSmaps} are complemented by the radial velocity $v_{rad}$ information, which already provides the line-of-sight information in an indirect way. A mapping of $v_{rad}$ to the missing $z$ coordinate is found in~\citet{Sawada_GC_face_on}, where it is given in the ($x,z$) projection (in our notation). Hence, the measured intensity of the radio emission $I(x,y,v_{rad})$ can be mapped to a full 3D cube $I(x,y,z)$ through a relation $v_{rad}(x,y) \leftrightarrow v_{rad}(x,z)$. The obtained cube gives an approximate picture of 3D gas distribution in the innermost ${\approx}400 \times 400 \times 100$~pc of the GC region.

Despite the limitations of this approach, based on this 3D cube one can quantify the biases in the calculations from the previous section. In particular, we can check whether the transition Eq.~\ref{eq::image_brightness}\textrightarrow~Eq.~\ref{eq::image_brightness_approx} still gives a valid estimate for a realistic gas density distribution. 

\begin{figure}
  \centering
  \includegraphics[width=\columnwidth]{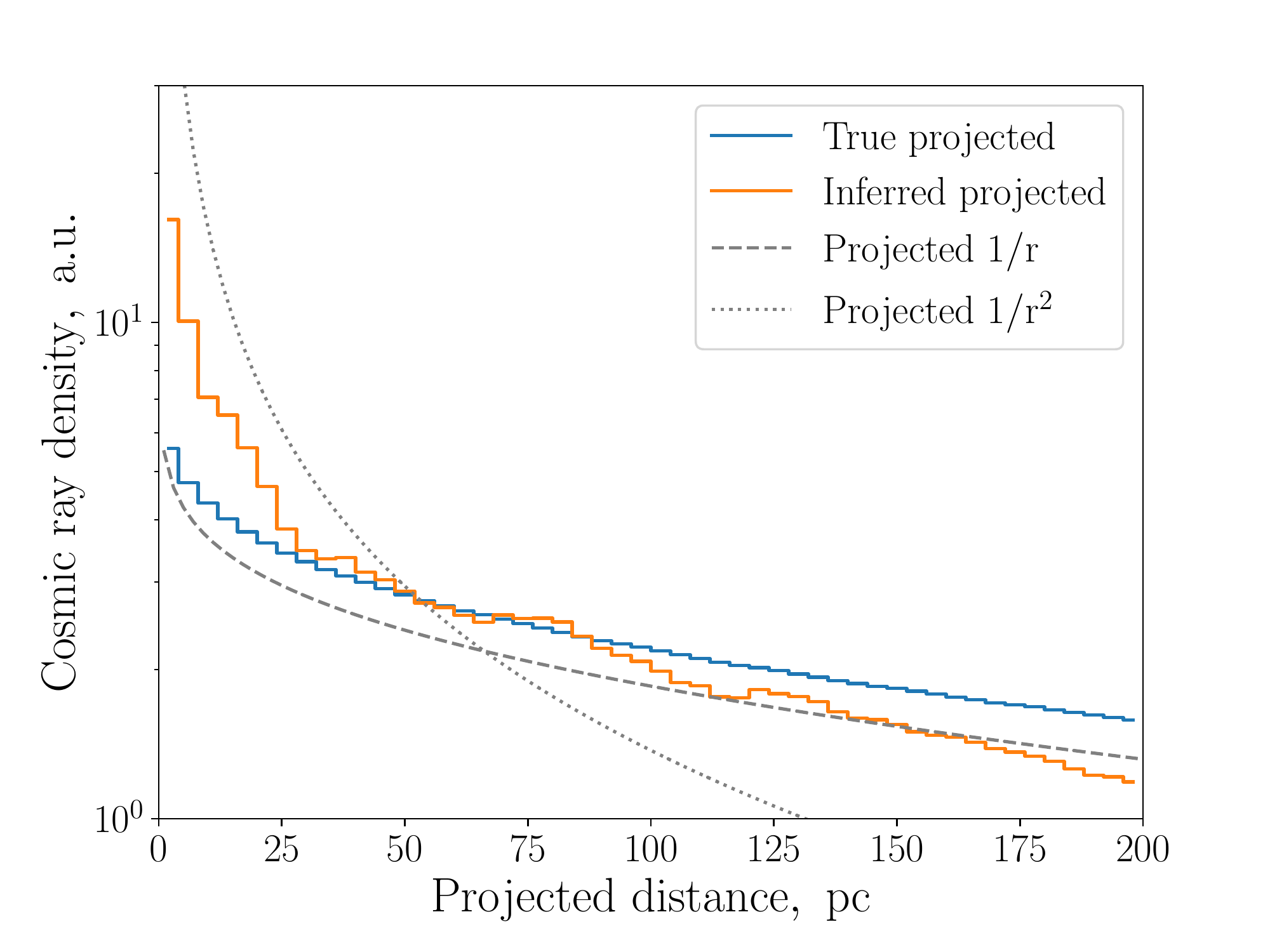}
  \caption{Projected cosmic-ray density as a function of the distance from the GC, computed using a 3D reconstruction of the gas distribution in the region. The underlining profile is assumed to have a $\rho_{CR} \propto r^{-1}$ shape. Blue line shows the true projection, whereas the orange one depicts the $P^\prime_{CR}(r) = S(r)/P_{gas}(r)$ estimate. Dashed and dotted grey lines represent the exact solutions for the $\alpha=\{1,2\}$ cases respectively, arbitrarily scaled to fit the image. The calculation was performed for the $\pm 10$~pc Galactic latitude slice along the Galactic plane.}
  \label{fig::CR_projected_density_issue}
\end{figure}

To verify this, we took the example case of a $\rho_{CR} \propto r^{-1}$ profile and computed the projected cosmic-ray density $P_{CR}(r)$ directly and using the $P^\prime_{CR}(r) = S(r)/P_{gas}(r)$ relation. The result of this comparison is shown in Fig.~\ref{fig::CR_projected_density_issue}, where the exact projected shapes for $\alpha=\{1,2\}$ profiles are also given.

\begin{figure}
  \centering
  \includegraphics[width=\columnwidth]{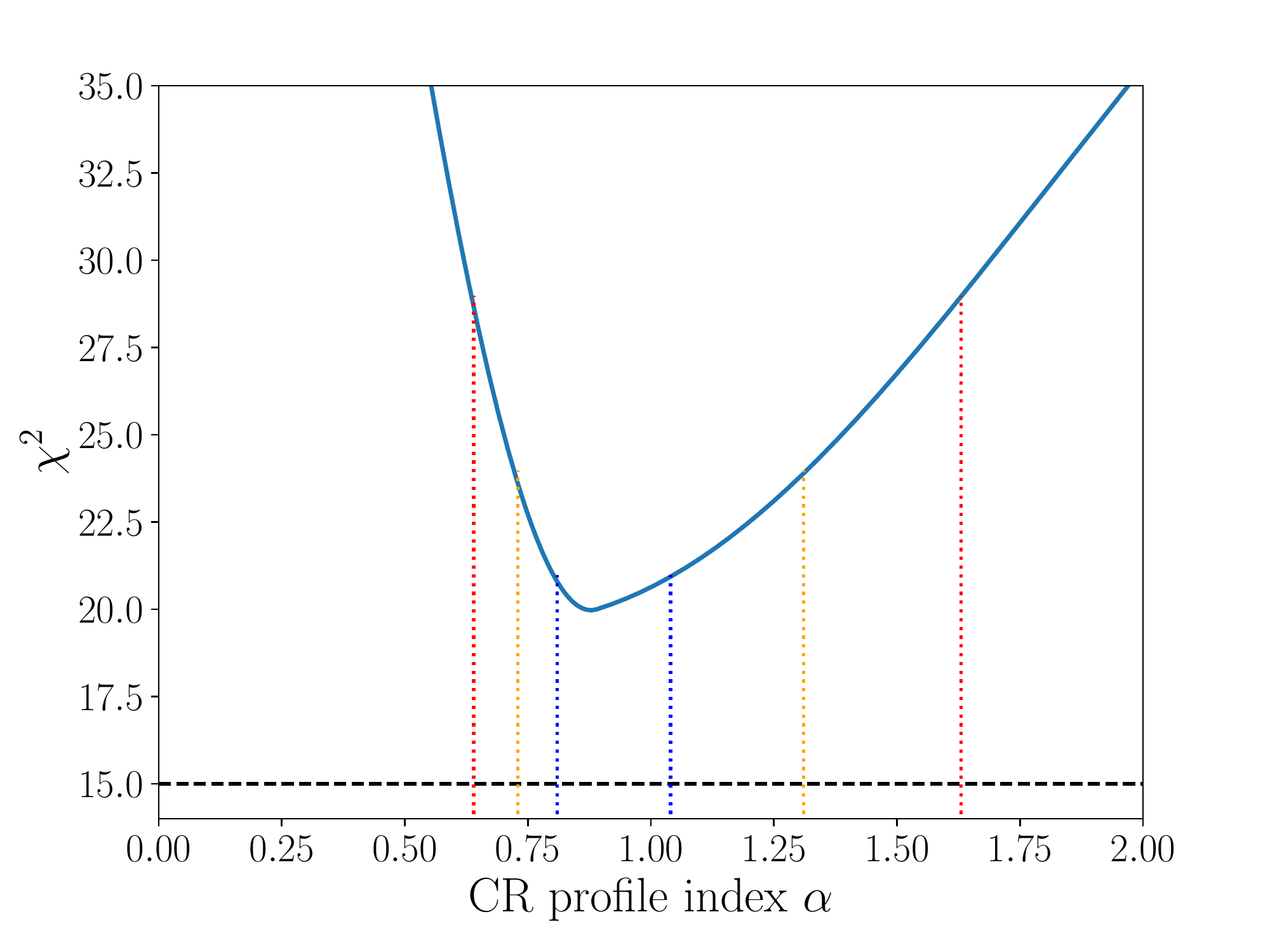}
  \caption{Scan of the cosmic-ray density profile power-law index $\alpha$ with respect to the data in Fig.~\ref{fig::CR_power_profile} and earlier measurements from~\citet{hess_collaboration_acceleration_2016}.
  The vertical dashed lines mark the derived 1 (blue), 2 (orange) and $3\sigma$ (red) confidence intervals.
  The horizontal dashed line represent the number of degrees of freedom in the fit.
  The scanned value, $\alpha$ corresponds to the true 3D cosmic-ray distribution and is projected on the plane of the sky using an assumed 3D gas map. See Sect.~\ref{sect::model_bias} for details.}
  \label{fig::CR_index_scan}
\end{figure}

It is clear from this figure that the inferred profile in the $\alpha=1$ case would resemble a steeper one with $\alpha^\prime \approx 1.5$. This way an $\alpha^\prime \approx 1.0-1.5$ measurement, presented in Sect.~\ref{sect::gal_plane_scan}, is likely to correspond to the true $\alpha \approx 1$ distribution -- accounting for this bias. In order to directly check this assumption, we have performed a fit of inferred profiles of a form $\rho_{CR} \propto r^{-\alpha} + const$ for $\alpha=[0;2]$ to the projected cosmic-ray densities, obtained here (Fig.~\ref{fig::CR_power_profile}) and earlier in~\citet{hess_collaboration_acceleration_2016}. The results of this test, shown in Fig.~\ref{fig::CR_index_scan}, indeed suggest that the true, de-projected cosmic-ray profile has a slope of $\alpha=0.88^{+0.16}_{-0.07}$ (at 68\% confidence level), consistent with the $\alpha=1$ assumption.

In a similar way we can also estimate the effect of the uncertain line-of-sight measurements on the derived value of $\alpha$. To reconstruct the effect we randomly shifted the positions of the cube grid cells in the $z$ direction and then regenerated the expected cosmic-ray profiles $P^\prime_{CR}(r)$. The random shifts were obtained by adding 10-40 Gaussian-distributed displacements with standard deviations from 50 to 200 pc at random positions within the cube. The amplitude of the Gaussians was fixed to 25 parsecs, while their width puts a coherence scale, not allowing the neighbouring bins to have very different shifts. The resulting shifts in the cube do not exceed ${\approx}\pm 50$~pc, roughly resembling the corresponding measurement uncertainties.

\begin{figure}
  \centering
  \includegraphics[width=\columnwidth]{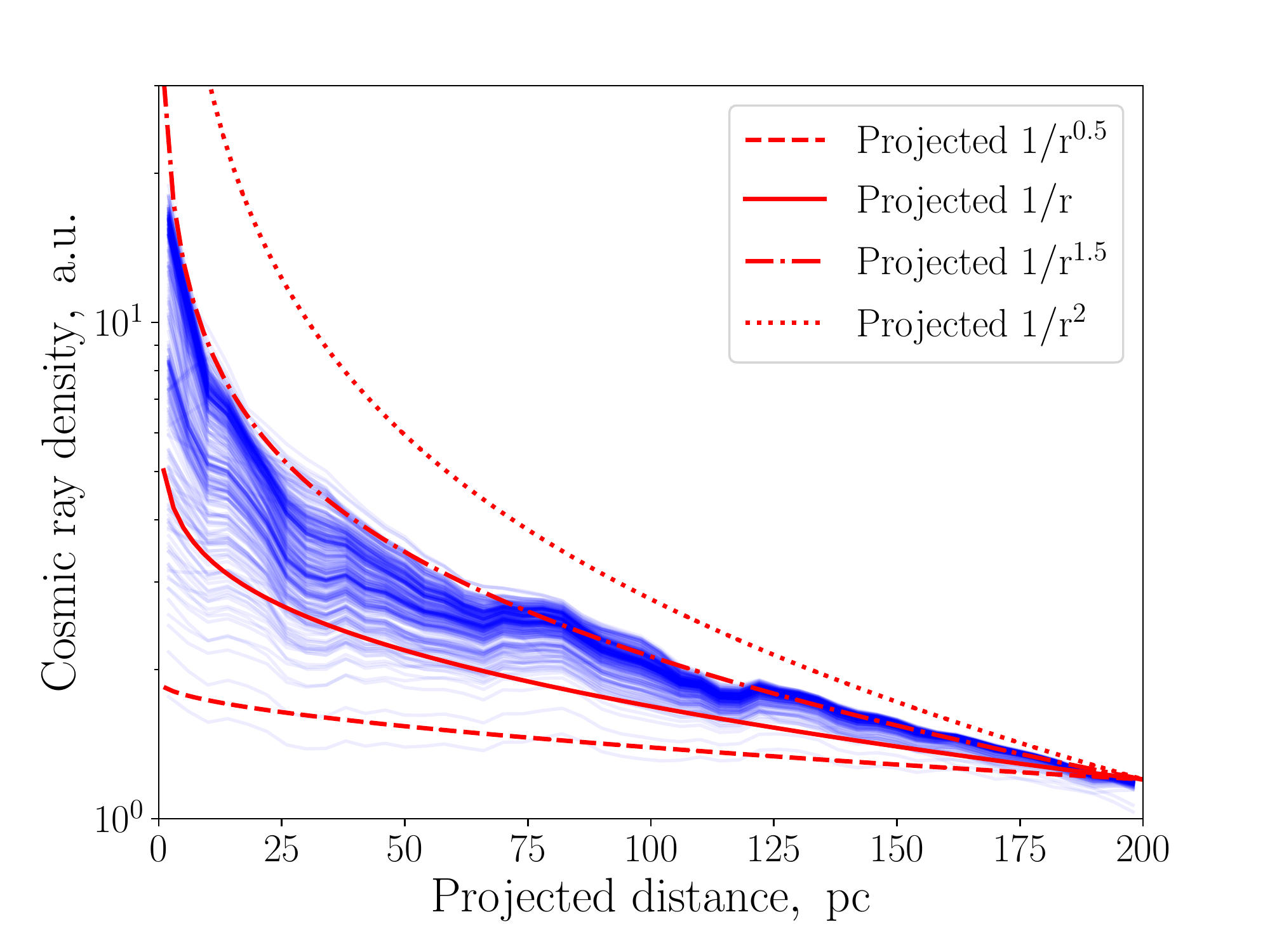}
  \caption{Projected cosmic-ray density as a function of the distance from the GC, computed using a 3D reconstruction of the gas distribution in the region. The profiles are inferred from the $P^\prime_{CR}(r) = S(r)/P_{gas}(r)$ estimate using the random line-of-sight shifts of the underlying gas distribution (see Sect.~\ref{sect::model_bias} for details). The true profile assumed in the simulation is $\rho_{CR} \approx r^{-1}$. Dashed and dotted red lines represent the exact solutions for the $\alpha=\{0.5, 1, 1.5, 2\}$ cases respectively, arbitrarily scaled to fit the image.}
  \label{fig::CR_projected_density_issue_scatter}
\end{figure}

The outcome of this calculation is shown in Fig.~\ref{fig::CR_projected_density_issue_scatter}. The comparison with the exact $\alpha=\{1,2\}$ profiles projections indicates that the uncertainty in the line-of-sight position measurements can dramatically change the shape of the measured cosmic-ray profile. Within our simulation setup, assuming an $\alpha=1$ cosmic-ray profile, we can generally conclude that a measurement of $\alpha^\prime \approx 1.5$ is equally likely.

It should also be noted that the main evidence for the centrally peaked cosmic-ray profile comes from the measurements in the central ${\approx}30-50$~pc, where the uncertainty on the gas content is the highest. Any unaccounted absorption of the molecular radio emission from such dense regions may lead to a general underestimation of the gas content in the direct vicinity of the GC (in addition to the line-of-sight uncertainty), and thereby to a more modest estimate of the cosmic-ray density there.


\section{Discussion}
\label{sect::discussion}

Based on deep VHE $\gamma$-ray observations using the MAGIC telescopes we can reconstruct the cosmic-ray distribution profile in the GC region. In this process we use the simplifications outlined in Sect.~\ref{sect::gal_plane_scan}, which result from our partial knowledge of the line-of-sight distribution of the target gas material. This implies that measurements of the cosmic-ray distribution in the GC vicinity can only be improved with more accurate gas observation in other (radio, infra-red etc) domains and are not limited by the sensitivity of current $\gamma$-ray instruments. Accordingly, even highly sensitive future CTA observations of this region, for an analysis like the one presented here, depend on higher-resolution molecular line measurements.

Our estimations described in Sec.~\ref{sect::model_bias} indicate that in the absence of such measurements the derived cosmic-ray profile may look sharper than it is in reality. At the same time, the scatter in the profiles, based on the quoted uncertainties in the line-of-sight gas clouds positions, may make any interpretation of the measured shape less reliable. As an example, the $\alpha=1$ cosmic-ray profile, simulated in Sect.~\ref{sect::model_bias}, may appear as $\alpha$ between 0.5 and 1.5 depending on the exact locations of the clouds (see Fig.~\ref{fig::CR_projected_density_issue_scatter}).

The existence of a centrally peaked cosmic-ray distribution around GC, revealed by the H.E.S.S. data~\citep{hess_collaboration_acceleration_2016} and confirmed here, however, seems reliable. The MAGIC data suggest an $\alpha = 1.2 \pm 0.3$ profile and therefore do not contradict the $\alpha=1$ scenario, in which CRs diffuse outwards from a central source, especially given the considerations outlined above.

This peaked cosmic-ray profile, together with a hard emission spectrum, was interpreted as a signature of PeV cosmic ray acceleration by SgrA*~\citep{hess_collaboration_acceleration_2016}. It should be noted here that alternative explanations for the enhanced density of CRs at GC exist, ranging from millisecond pulsars~\citep{guepin_pevatron_2018} to dark matter~\citep{lacroix_connecting_2016}. It has also been shown by \citet{gaggero17} that it is possible to consistently model the diffuse $\gamma$-ray emission from GC in the Fermi and VHE emission energy range with particles from the Galactic CR sea, nearly without the need of an addition central CR source. The authors also state that in the presence of such a CR sea, the maximum energy of any excess from a central source is even less certain.

MAGIC measurements of the diffuse $\gamma$-ray spectrum over a larger region of the central molecular zone, ${\approx}150$~pc in width, favor, on a ${\approx}2\sigma$ level, a cut-off in the $\gamma$-ray spectrum over a pure power law. The $1\sigma$ confidence range for the cut-off energy spans from $10$~TeV to $80$~TeV (see Sect.~\ref{sect::spectra}). This corresponds to proton energies of ${\approx}0.1-1$~PeV, which means that the data are still marginally compatible with the PeVatron scenario. 

\section{Conclusions}
\label{sect::conclusions}

The presence and proximity of a central super-massive black hole make the GC region a unique laboratory for studying comic-ray acceleration near black holes. The above presented MAGIC observations confirm that the cosmic-ray density profile in the GC vicinity is peaked towards the position of the central SMBH of our Galaxy, similar to what is pointed out by \citet{hess_collaboration_acceleration_2016}. This is consistent with the idea of activity of the latter as a particle accelerator in the recent past or, possibly, even nowadays. The MAGIC data also show that the spectrum of this diffuse gamma-ray emission is hard ($\Gamma \approx 2$) and reaches energies of several ten TeV. Our analysis, however, also revealed a hint for a spectral turnover at around those energies at the $2 \sigma$ level.

In order to gain further insight to particle acceleration and diffusion in the GC region, including the PeVatron topic, even deeper observations of the GC region by current and next-generation instruments are necessary. The focus should be on energies around and above 10~TeV, where current data sets are still statistics limited. Our study also indicates that such detailed measurements will be very sensitive to the line-of-sight gas distribution assumed and, as such, would require a deeper and higher resolution radio survey of the central 200~pc of our Galaxy.

\begin{acknowledgements}
We would like to thank the Instituto de Astrof\'{\i}sica de Canarias for the excellent working conditions at the Observatorio del Roque de los Muchachos in La Palma. We also thank Dario Grasso for the valuable comments. The financial support of the German BMBF and MPG, the Italian INFN and INAF, the Swiss National Fund SNF, the ERDF under the Spanish MINECO (FPA2015-69818-P, FPA2012-36668, FPA2015-68378-P, FPA2015-69210-C6-2-R, FPA2015-69210-C6-4-R, FPA2015-69210-C6-6-R, AYA2015-71042-P, AYA2016-76012-C3-1-P, ESP2015-71662-C2-2-P, FPA2017‐90566‐REDC), the Indian Department of Atomic Energy, the Japanese JSPS and MEXT, the Bulgarian Ministry of Education and Science, National RI Roadmap Project DO1-153/28.08.2018 and the Academy of Finland grant nr. 320045 is gratefully acknowledged. This work was also supported by the Spanish Centro de Excelencia ``Severo Ochoa'' SEV-2016-0588 and SEV-2015-0548, and Unidad de Excelencia ``Mar\'{\i}a de Maeztu'' MDM-2014-0369, by the Croatian Science Foundation (HrZZ) Project IP-2016-06-9782 and the University of Rijeka Project 13.12.1.3.02, by the DFG Collaborative Research Centers SFB823/C4 and SFB876/C3, the Polish National Research Centre grant UMO-2016/22/M/ST9/00382 and by the Brazilian MCTIC, CNPq and FAPERJ.
\end{acknowledgements}


\bibliographystyle{aa}
\bibliography{GalDiffuse_references}

\begin{thebibliography}{28}
\expandafter\ifx\csname natexlab\endcsname\relax\def\natexlab#1{#1}\fi

\bibitem[{Abdo {et~al.}(2010)Abdo, Ackermann, Ajello, Atwood, Baldini, Ballet,
  Barbiellini, Bastieri, Baughman, Bechtol, Bellazzini, Berenji, Blandford,
  Bloom, Bonamente, Borgland, Bregeon, Brez, Brigida, Bruel, Burnett, Buson,
  Caliandro, Cameron, Caraveo, Casandjian, Cavazzuti, Cecchi, Çelik, Charles,
  Chekhtman, Cheung, Chiang, Ciprini, Claus, Cohen-Tanugi, Cominsky, Conrad,
  Cutini, Dermer, de~Angelis, de~Palma, Digel, Di~Bernardo, e~Silva, Drell,
  Drlica-Wagner, Dubois, Dumora, Farnier, Favuzzi, Fegan, Focke, Fortin,
  Frailis, Fukazawa, Funk, Fusco, Gaggero, Gargano, Gasparrini, Gehrels,
  Germani, Giebels, Giglietto, Giommi, Giordano, Glanzman, Godfrey, Grenier,
  Grondin, Grove, Guillemot, Guiriec, Gustafsson, Hanabata, Harding, Hayashida,
  Hughes, Itoh, Jackson, Jóhannesson, Johnson, Johnson, Johnson, Johnson,
  Kamae, Katagiri, Kataoka, Kawai, Kerr, Knödlseder, Kocian, Kuehn, Kuss,
  Lande, Latronico, Lemoine-Goumard, Longo, Loparco, Lott, Lovellette, Lubrano,
  Madejski, Makeev, Mazziotta, McConville, McEnery, Meurer, Michelson,
  Mitthumsiri, Mizuno, Moiseev, Monte, Monzani, Morselli, Moskalenko, Murgia,
  Nolan, Norris, Nuss, Ohsugi, Omodei, Orlando, Ormes, Paneque, Panetta,
  Parent, Pelassa, Pepe, Pesce-Rollins, Piron, Porter, Rainò, Rando, Razzano,
  Reimer, Reimer, Reposeur, Ritz, Rochester, Rodriguez, Roth, Ryde,
  Sadrozinski, Sanchez, Sander, Parkinson, Scargle, Sellerholm, Sgrò, Shaw,
  Siskind, Smith, Smith, Spandre, Spinelli, Starck, Strickman, Strong, Suson,
  Tajima, Takahashi, Takahashi, Tanaka, Thayer, Thayer, Thompson, Tibaldo,
  Torres, Tosti, Tramacere, Uchiyama, Usher, Vasileiou, Vilchez, Vitale, Waite,
  Wang, Winer, Wood, Ylinen, \& Ziegler}]{abdo_spectrum_2010}
Abdo, A.~A., Ackermann, M., Ajello, M., {et~al.} 2010, Physical Review Letters,
  104

\bibitem[{{Aharonian} {et~al.}(2006{\natexlab{b}}){Aharonian}, {Akhperjanian},
  {Bazer-Bachi}, {Beilicke}, {Benbow}, {Berge}, {Bernl{\"o}hr}, {Boisson},
  {Bolz}, {Borrel}, {Braun}, {Breitling}, {Brown}, {Chadwick}, {Chounet},
  {Cornils}, {Costamante}, {Degrange}, {Dickinson}, {Djannati-Ata{\"i}},
  {Drury}, {Dubus}, {Emmanoulopoulos}, {Espigat}, {Feinstein}, {Fontaine},
  {Fuchs}, {Funk}, {Gallant}, {Giebels}, {Gillessen}, {Glicenstein}, {Goret},
  {Hadjichristidis}, {Hauser}, {Hauser}, {Heinzelmann}, {Henri}, {Hermann},
  {Hinton}, {Hofmann}, {Holleran}, {Horns}, {Jacholkowska}, {de Jager},
  {Kh{\'e}lifi}, {Klages}, {Komin}, {Konopelko}, {Latham}, {Le Gallou},
  {Lemi{\`e}re}, {Lemoine-Goumard}, {Leroy}, {Lohse}, {Marcowith}, {Martin},
  {Martineau-Huynh}, {Masterson}, {McComb}, {de Naurois}, {Nolan}, {Noutsos},
  {Orford}, {Osborne}, {Ouchrif}, {Panter}, {Pelletier}, {Pita},
  {P{\"u}hlhofer}, {Punch}, {Raubenheimer}, {Raue}, {Raux}, {Rayner}, {Reimer},
  {Reimer}, {Ripken}, {Rob}, {Rolland}, {Rowell}, {Sahakian}, {Saug{\'e}},
  {Schlenker}, {Schlickeiser}, {Schuster}, {Schwanke}, {Siewert}, {Sol},
  {Spangler}, {Steenkamp}, {Stegmann}, {Tavernet}, {Terrier}, {Th{\'e}oret},
  {Tluczykont}, {van Eldik}, {Vasileiadis}, {Venter}, {Vincent}, {V{\"o}lk}, \&
  {Wagner}}]{HESS_VHE_GC_ridge}
{Aharonian}, F., {Akhperjanian}, A.~G., {Bazer-Bachi}, A.~R., {et~al.}
  2006{\natexlab{b}}, \nat, 439, 695

\bibitem[{{Aharonian} {et~al.}(2006{\natexlab{a}}){Aharonian}, {Akhperjanian},
  {Bazer-Bachi}, {Beilicke}, {Benbow}, {Berge}, {Bernl{\"o}hr}, {Boisson},
  {Bolz}, {Borrel}, {Braun}, {Breitling}, {Brown}, {Chadwick}, {Chounet},
  {Cornils}, {Costamante}, {Degrange}, {Dickinson}, {Djannati-Ata{\"i}},
  {Drury}, {Dubus}, {Emmanoulopoulos}, {Espigat}, {Feinstein}, {Fontaine},
  {Fuchs}, {Funk}, {Gallant}, {Giebels}, {Gillessen}, {Glicenstein}, {Goret},
  {Hadjichristidis}, {Hauser}, {Heinzelmann}, {Henri}, {Hermann}, {Hinton},
  {Hofmann}, {Holleran}, {Horns}, {Jacholkowska}, {de Jager}, {Kh{\'e}lifi},
  {Komin}, {Konopelko}, {Latham}, {Le Gallou}, {Lemi{\`e}re},
  {Lemoine-Goumard}, {Leroy}, {Lohse}, {Martin}, {Martineau-Huynh},
  {Marcowith}, {Masterson}, {McComb}, {de Naurois}, {Nolan}, {Noutsos},
  {Orford}, {Osborne}, {Ouchrif}, {Panter}, {Pelletier}, {Pita},
  {P{\"u}hlhofer}, {Punch}, {Raubenheimer}, {Raue}, {Raux}, {Rayner}, {Reimer},
  {Reimer}, {Ripken}, {Rob}, {Rolland}, {Rowell}, {Sahakian}, {Saug{\'e}},
  {Schlenker}, {Schlickeiser}, {Schuster}, {Schwanke}, {Siewert}, {Sol},
  {Spangler}, {Steenkamp}, {Stegmann}, {Tavernet}, {Terrier}, {Th{\'e}oret},
  {Tluczykont}, {Vasileiadis}, {Venter}, {Vincent}, {V{\"o}lk}, \&
  {Wagner}}]{aharonian2006hess}
{Aharonian}, F., {Akhperjanian}, A.~G., {Bazer-Bachi}, A.~R., {et~al.}
  2006{\natexlab{a}}, \apj, 636, 777

\bibitem[{{Ahnen} {et~al.}(2017{\natexlab{a}}){Ahnen}, {Ansoldi}, {Antonelli},
  {et~al.}}]{MAGIC_GC_2017}
{Ahnen}, M.~L., {Ansoldi}, S., {Antonelli}, L.~A., {et~al.} 2017{\natexlab{a}},
  \aap, 601, A33

\bibitem[{{Ahnen} {et~al.}(2017{\natexlab{b}}){Ahnen}, {Ansoldi}, {Antonelli},
  {et~al.}}]{ahnen_performance_2017}
{Ahnen}, M.~L., {Ansoldi}, S., {Antonelli}, L.~A., {et~al.} 2017{\natexlab{b}},
  Astroparticle Physics, 94, 29

\bibitem[{{Albert} {et~al.}(2008){Albert}, {Aliu}, {Anderhub}, {Antoranz},
  {Armada}, {Asensio}, {Baixeras}, {Barrio}, {Bartko}, {Bastieri}, {Becker},
  {Bednarek}, {Berger}, {Bigongiari}, {Biland}, {Bock}, {Bordas},
  {Bosch-Ramon}, {Bretz}, {Britvitch}, {Camara}, {Carmona}, {Chilingarian},
  {Ciprini}, {Coarasa}, {Commichau}, {Contreras}, {Cortina}, {Costado},
  {Curtef}, {Danielyan}, {Dazzi}, {de Angelis}, {Delgado}, {de Los Reyes}, {de
  Lotto}, {Domingo-Santamar{\'{\i}}a}, {Dorner}, {Doro}, {Errando},
  {Fagiolini}, {Ferenc}, {Fern{\'a}ndez}, {Firpo}, {Flix}, {Fonseca}, {Font},
  {Fuchs}, {Galante}, {Garc{\'{\i}}a-L{\'o}pez}, {Garczarczyk}, {Gaug},
  {Giller}, {Goebel}, {Hakobyan}, {Hayashida}, {Hengstebeck}, {Herrero},
  {H{\"o}hne}, {Hose}, {Huber}, {Hsu}, {Jacon}, {Jogler}, {Kosyra}, {Kranich},
  {Kritzer}, {Laille}, {Lindfors}, {Lombardi}, {Longo}, {L{\'o}pez},
  {L{\'o}pez}, {Lorenz}, {Majumdar}, {Maneva}, {Mannheim}, {Mariotti},
  {Mart{\'{\i}}nez}, {Mazin}, {Merck}, {Meucci}, {Meyer}, {Miranda},
  {Mirzoyan}, {Mizobuchi}, {Moralejo}, {Nieto}, {Nilsson}, {Ninkovic},
  {O{\~n}a-Wilhelmi}, {Otte}, {Oya}, {Panniello}, {Paoletti}, {Paredes},
  {Pasanen}, {Pascoli}, {Pauss}, {Pegna}, {Persic}, {Peruzzo}, {Piccioli},
  {Puchades}, {Prandini}, {Raymers}, {Rhode}, {Rib{\'o}}, {Rico}, {Rissi},
  {Robert}, {R{\"u}gamer}, {Saggion}, {Saito}, {S{\'a}nchez}, {Sartori},
  {Scalzotto}, {Scapin}, {Schmitt}, {Schweizer}, {Shayduk}, {Shinozaki},
  {Shore}, {Sidro}, {Sillanp{\"a}{\"a}}, {Sobczynska}, {Spanier}, {Stamerra},
  {Stark}, {Takalo}, {Temnikov}, {Tescaro}, {Teshima}, {Torres}, {Turini},
  {Vankov}, {Venturini}, {Vitale}, {Wagner}, {Wibig}, {Wittek}, {Zandanel},
  {Zanin}, \& {Zapatero}}]{Magic_RF}
{Albert}, J., {Aliu}, E., {Anderhub}, H., {et~al.} 2008, Nuclear Instruments
  and Methods in Physics Research A, 588, 424

\bibitem[{{Aleksi{\'c}} {et~al.}(2016){Aleksi{\'c}}, {Ansoldi}, {Antonelli},
  {Antoranz}, {Babic}, {Bangale}, {Barcel{\'o}}, {Barrio}, {Becerra
  Gonz{\'a}lez}, {Bednarek}, {Bernardini}, {Biasuzzi}, {Biland}, {Bitossi},
  {Blanch}, {Bonnefoy}, {Bonnoli}, {Borracci}, {Bretz}, {Carmona}, {Carosi},
  {Cecchi}, {Colin}, {Colombo}, {Contreras}, {Corti}, {Cortina}, {Covino}, {Da
  Vela}, {Dazzi}, {De Angelis}, {De Caneva}, {De Lotto}, {de O{\~n}a Wilhelmi},
  {Delgado Mendez}, {Dettlaff}, {Dominis Prester}, {Dorner}, {Doro}, {Einecke},
  {Eisenacher}, {Elsaesser}, {Fidalgo}, {Fink}, {Fonseca}, {Font}, {Frantzen},
  {Fruck}, {Galindo}, {Garc{\'{\i}}a L{\'o}pez}, {Garczarczyk}, {Garrido
  Terrats}, {Gaug}, {Giavitto}, {Godinovi{\'c}}, {Gonz{\'a}lez Mu{\~n}oz},
  {Gozzini}, {Haberer}, {Hadasch}, {Hanabata}, {Hayashida}, {Herrera},
  {Hildebrand}, {Hose}, {Hrupec}, {Idec}, {Illa}, {Kadenius}, {Kellermann},
  {Knoetig}, {Kodani}, {Konno}, {Krause}, {Kubo}, {Kushida}, {La Barbera},
  {Lelas}, {Lemus}, {Lewandowska}, {Lindfors}, {Lombardi}, {Longo},
  {L{\'o}pez}, {L{\'o}pez-Coto}, {L{\'o}pez-Oramas}, {Lorca}, {Lorenz},
  {Lozano}, {Makariev}, {Mallot}, {Maneva}, {Mankuzhiyil}, {Mannheim},
  {Maraschi}, {Marcote}, {Mariotti}, {Mart{\'{\i}}nez}, {Mazin}, {Menzel},
  {Miranda}, {Mirzoyan}, {Moralejo}, {Munar-Adrover}, {Nakajima}, {Negrello},
  {Neustroev}, {Niedzwiecki}, {Nilsson}, {Nishijima}, {Noda}, {Orito},
  {Overkemping}, {Paiano}, {Palatiello}, {Paneque}, {Paoletti}, {Paredes},
  {Paredes-Fortuny}, {Persic}, {Poutanen}, {Prada Moroni}, {Prandini},
  {Puljak}, {Reinthal}, {Rhode}, {Rib{\'o}}, {Rico}, {Rodriguez Garcia},
  {R{\"u}gamer}, {Saito}, {Saito}, {Satalecka}, {Scalzotto}, {Scapin},
  {Schultz}, {Schlammer}, {Schmidl}, {Schweizer}, {Shore}, {Sillanp{\"a}{\"a}},
  {Sitarek}, {Snidaric}, {Sobczynska}, {Spanier}, {Stamerra}, {Steinbring},
  {Storz}, {Strzys}, {Takalo}, {Takami}, {Tavecchio}, {Tejedor}, {Temnikov},
  {Terzi{\'c}}, {Tescaro}, {Teshima}, {Thaele}, {Tibolla}, {Torres}, {Toyama},
  {Treves}, {Vogler}, {Wetteskind}, {Will}, \& {Zanin}}]{Magic_performanceII}
{Aleksi{\'c}}, J., {Ansoldi}, S., {Antonelli}, L.~A., {et~al.} 2016,
  Astroparticle Physics, 72, 76

\bibitem[{{Archer} {et~al.}(2016){Archer}, {Benbow}, {Bird}, {Buchovecky},
  {Buckley}, {Bugaev}, {Byrum}, {Cardenzana}, {Cerruti}, {Chen}, {Ciupik},
  {Collins-Hughes}, {Connolly}, {Eisch}, {Falcone}, {Feng}, {Finley},
  {Fleischhack}, {Flinders}, {Fortson}, {Furniss}, {Gillanders}, {Griffin},
  {Grube}, {Gyuk}, {Hakansson}, {Hanna}, {Holder}, {Humensky}, {Hutten},
  {Johnson}, {Kaaret}, {Kar}, {Kelley-Hoskins}, {Kertzman}, {Kieda}, {Krause},
  {Krennrich}, {Kumar}, {Lang}, {McArthur}, {McCann}, {Meagher}, {Millis},
  {Moriarty}, {Mukherjee}, {Nieto}, {Ong}, {Park}, {Pelassa}, {Pohl}, {Popkow},
  {Pueschel}, {Quinn}, {Ragan}, {Ratliff}, {Reynolds}, {Richards}, {Roache},
  {Rousselle}, {Santander}, {Sembroski}, {Shahinyan}, {Smith}, {Staszak},
  {Telezhinsky}, {Tucci}, {Tyler}, {Vassiliev}, {Wakely}, {Weiner},
  {Weinstein}, {Wilhelm}, {Williams}, {Zitzer}, \&
  {Yusef-Zadeh}}]{2016VERITASGC_Smith}
{Archer}, A., {Benbow}, W., {Bird}, R., {et~al.} 2016, ArXiv e-prints:
  1602.08522

\bibitem[{Collaboration \& Aharonian(2006)}]{hess_Galactic_ridge}
Collaboration, T. H. E. S.~S. \& Aharonian, F.~A. 2006, Nature, 439, 695,
  arXiv: astro-ph/0603021

\bibitem[{{Ferri{\`e}re} {et~al.}(2007){Ferri{\`e}re}, {Gillard}, \&
  {Jean}}]{ferriere07}
{Ferri{\`e}re}, K., {Gillard}, W., \& {Jean}, P. 2007, \aap, 467, 611

\bibitem[{{Fomin} {et~al.}(1994){Fomin}, {Stepanian}, {Lamb}, {Lewis}, {Punch},
  \& {Weekes}}]{1994Fomin_wobble}
{Fomin}, V.~P., {Stepanian}, A.~A., {Lamb}, R.~C., {et~al.} 1994, Astroparticle
  Physics, 2, 137

\bibitem[{{Foreman-Mackey} {et~al.}(2013){Foreman-Mackey}, {Hogg}, {Lang}, \&
  {Goodman}}]{emcee}
{Foreman-Mackey}, D., {Hogg}, D.~W., {Lang}, D., \& {Goodman}, J. 2013, \pasp,
  125, 306

\bibitem[{Fruck {et~al.}(2014)Fruck, Gaug, Zanin, Dorner, Garrido, Mirzoyan,
  Font, \& Collaboration}]{fruck_novel_2014}
Fruck, C., Gaug, M., Zanin, R., {et~al.} 2014, arXiv:1403.3591 [astro-ph],
  arXiv: 1403.3591

\bibitem[{{Gaggero} {et~al.}(2017){Gaggero}, {Grasso}, {Marinelli}, {Taoso}, \&
  {Urbano}}]{gaggero17}
{Gaggero}, D., {Grasso}, D., {Marinelli}, A., {Taoso}, M., \& {Urbano}, A.
  2017, Physical Review Letters, 119, 031101

\bibitem[{Gaug {et~al.}(2014)Gaug, Blanch, Dorner, Doro, Font, Fruck,
  Garczarczyk, Garrido, Hrupec, Hose, López-Oramas, Maneva, Martinez,
  Mirzoyan, Temnikov, \& Zanin}]{gaug_atmospheric_2014}
Gaug, M., Blanch, O., Dorner, D., {et~al.} 2014, arXiv:1403.5083 [astro-ph],
  arXiv: 1403.5083

\bibitem[{Guépin {et~al.}(2018)Guépin, Rinchiuso, Kotera, Moulin, Pierog, \&
  Silk}]{guepin_pevatron_2018}
Guépin, C., Rinchiuso, L., Kotera, K., {et~al.} 2018, Journal of Cosmology and
  Astroparticle Physics, 2018, 042

\bibitem[{{H.E.S.S.} {et~al.}(2018){H.E.S.S.}, {Abdalla, H.}, {Abramowski, A.},
  {Aharonian, F.}, {Benkhali, F. Ait}, {Akhperjanian, A. G.}, {Andersson, T.},
  {Ang\"uner, E. O.}, {Arakawa, M.}, {Arrieta, M.}, {Aubert, P.}, {Backes, M.},
  {Balzer, A.}, {Barnard, M.}, {Becherini, Y.}, {Tjus, J. Becker}, {Berge, D.},
  {Bernhard, S.}, {Bernl\"ohr, K.}, {Blackwell, R.}, {B\"ottcher, M.},
  {Boisson, C.}, {Bolmont, J.}, {Bonnefoy, S.}, {Bordas, P.}, {Bregeon, J.},
  {Brun, F.}, {Brun, P.}, {Bryan, M.}, {B\"uchele, M.}, {Bulik, T.}, {Capasso,
  M.}, {Carr, J.}, {Casanova, S.}, {Cerruti, M.}, {Chakraborty, N.}, {G.
  Chaves, R. C.}, {Chen, A.}, {Chevalier, J.}, {Coffaro, M.}, {Colafrancesco,
  S.}, {Cologna, G.}, {Condon, B.}, {Conrad, J.}, {Cui, Y.}, {Davids, I. D.},
  {Decock, J.}, {Degrange, B.}, {Deil, C.}, {Devin, J.}, {deWilt, P.}, {Dirson,
  L.}, {Djannati-Ata\"{\i}, A.}, {Domainko, W.}, {Donath, A.}, {Drury, L.
  O\'{}C.}, {Dutson, K.}, {Dyks, J.}, {Edwards, T.}, {Egberts, K.}, {Eger, P.},
  {Ernenwein, J.-P.}, {Eschbach, S.}, {Farnier, C.}, {Fegan, S.}, {Fernandes,
  M. V.}, {Fiasson, A.}, {Fontaine, G.}, {F\"orster, A.}, {Funk, S.},
  {F\"u\ss{}ling, M.}, {Gabici, S.}, {Gallant, Y. A.}, {Garrigoux, T.},
  {Giavitto, G.}, {Giebels, B.}, {Glicenstein, J. F.}, {Gottschall, D.},
  {Goyal, A.}, {Grondin, M.-H.}, {Hahn, J.}, {Haupt, M.}, {Hawkes, J.},
  {Heinzelmann, G.}, {Henri, G.}, {Hermann, G.}, {Hinton, J. A.}, {Hofmann,
  W.}, {Hoischen, C.}, {Holch, T. L.}, {Holler, M.}, {Horns, D.}, {Ivascenko,
  A.}, {Iwasaki, H.}, {Jacholkowska, A.}, {Jamrozy, M.}, {Janiak, M.},
  {Jankowsky, D.}, {Jankowsky, F.}, {Jingo, M.}, {Jogler, T.}, {Jouvin, L.},
  {Jung-Richardt, I.}, {Kastendieck, M. A.}, {Katarzy\'{}nski, K.},
  {Katsuragawa, M.}, {Katz, U.}, {Kerszberg, D.}, {Khangulyan, D.}, {Kh\'elifi,
  B.}, {King, J.}, {Klepser, S.}, {Klochkov, D.}, {Klu\'{}zniak, W.},
  {Kolitzus, D.}, {Komin, Nu.}, {Kosack, K.}, {Krakau, S.}, {Kraus, M.},
  {Kr\"uger, P. P.}, {Laffon, H.}, {Lamanna, G.}, {Lau, J.}, {Lees, J.-P.},
  {Lefaucheur, J.}, {Lefranc, V.}, {Lemi\`ere, A.}, {Lemoine-Goumard, M.},
  {Lenain, J.-P.}, {Leser, E.}, {Lohse, T.}, {Lorentz, M.}, {Liu, R.},
  {L\'opez-Coto, R.}, {Lypova, I.}, {Marandon, V.}, {Marcowith, A.}, {Mariaud,
  C.}, {Marx, R.}, {Maurin, G.}, {Maxted, N.}, {Mayer, M.}, {Meintjes, P. J.},
  {Meyer, M.}, {W. Mitchell, A. M.}, {Moderski, R.}, {Mohamed, M.}, {Mohrmann,
  L.}, {Mor\aa{}, K.}, {Moulin, E.}, {Murach, T.}, {Nakashima, S.}, {Naurois,
  M. de}, {Niederwanger, F.}, {Niemiec, J.}, {Oakes, L.}, {O\'{}Brien, P.},
  {Odaka, H.}, {Ohm, S.}, {Ostrowski, M.}, {Oya, I.}, {Padovani, M.}, {Panter,
  M.}, {Parsons, R. D.}, {Pekeur, N. W.}, {Pelletier, G.}, {Perennes, C.},
  {Petrucci, P.-O.}, {Peyaud, B.}, {Piel, Q.}, {Pita, S.}, {Poon, H.},
  {Prokhorov, D.}, {Prokoph, H.}, {P\"uhlhofer, G.}, {Punch, M.}, {Quirrenbach,
  A.}, {Raab, S.}, {Rauth, R.}, {Reimer, A.}, {Reimer, O.}, {Renaud, M.}, {los
  Reyes, R. de}, {Richter, S.}, {Rieger, F.}, {Romoli, C.}, {Rowell, G.},
  {Rudak, B.}, {Rulten, C. B.}, {Sahakian, V.}, {Saito, S.}, {Salek, D.},
  {Sanchez, D. A.}, {Santangelo, A.}, {Sasaki, M.}, {Schlickeiser, R.},
  {Sch\"ussler, F.}, {Schulz, A.}, {Schwanke, U.}, {Schwemmer, S.},
  {Seglar-Arroyo, M.}, {Settimo, M.}, {Seyffert, A. S.}, {Shafi, N.}, {Shilon,
  I.}, {Simoni, R.}, {Sol, H.}, {Spanier, F.}, {Spengler, G.}, {Spies, F.},
  {Stawarz, L.}, {Steenkamp, R.}, {Stegmann, C.}, {Stycz, K.}, {Sushch, I.},
  {Takahashi, T.}, {Tavernet, J.-P.}, {Tavernier, T.}, {Taylor, A. M.},
  {Terrier, R.}, {Tibaldo, L.}, {Tiziani, D.}, {Tluczykont, M.}, {Trichard,
  C.}, {Tsuji, N.}, {Tuffs, R.}, {Uchiyama, Y.}, {van der Walt, D. J.}, {Eldik,
  C. van}, {Rensburg, C. van}, {Soelen, B. van}, {Vasileiadis, G.}, {Veh, J.},
  {Venter, C.}, {Viana, A.}, {Vincent, P.}, {Vink, J.}, {Voisin, F.}, {V\"olk,
  H. J.}, {Vuillaume, T.}, {Wadiasingh, Z.}, {Wagner, S. J.}, {Wagner, P.},
  {Wagner, R. M.}, {White, R.}, {Wierzcholska, A.}, {Willmann, P.},
  {W\"ornlein, A.}, {Wouters, D.}, {Yang, R.}, {Zaborov, D.}, {Zacharias, M.},
  {Zanin, R.}, {Zdziarski, A. A.}, {Zech, A.}, {Zefi, F.}, {Ziegler, A.}, \&
  {Zywucka, N.}}]{HESS_GalDiffuse_2018}
{H.E.S.S.}, {Abdalla, H.}, {Abramowski, A.}, {et~al.} 2018, A\&A, 612, A9

\bibitem[{{H.E.S.S.} {et~al.}(2016){H.E.S.S.}, {Abramowski}, {Aharonian},
  {Benkhali}, {Akhperjanian}, {Ang{\"u}ner}, {Backes}, {Balzer}, {Becherini},
  {Tjus}, \& et~al.}]{hess_collaboration_acceleration_2016}
{H.E.S.S.}, {Abramowski}, A., {Aharonian}, F., {et~al.} 2016, \nat, 531, 476

\bibitem[{Hunter {et~al.}(1997)Hunter, Bertsch, Catelli, Dame, Digel, Dingus,
  Esposito, Fichtel, Hartman, Kanbach, Kniffen, Lin, Mayer-Hasselwander,
  Michelson, Montigny, Mukherjee, Nolan, Schneid, Sreekumar, Thaddeus, \&
  Thompson}]{hunter_egret_1997}
Hunter, S.~D., Bertsch, D.~L., Catelli, J.~R., {et~al.} 1997, The Astrophysical
  Journal, 481, 205

\bibitem[{Lacroix {et~al.}(2016)Lacroix, Silk, Moulin, \&
  Bœhm}]{lacroix_connecting_2016}
Lacroix, T., Silk, J., Moulin, E., \& Bœhm, C. 2016, Physical Review D, 94,
  123008

\bibitem[{{Nakanishi} \& {Sofue}(2003)}]{nakanish03}
{Nakanishi}, H. \& {Sofue}, Y. 2003, \pasj, 55, 191

\bibitem[{Oka {et~al.}(1998)Oka, Hasegawa, Sato, Tsuboi, \&
  Miyazaki}]{oka_largescale_1998}
Oka, T., Hasegawa, T., Sato, F., Tsuboi, M., \& Miyazaki, A. 1998, The
  Astrophysical Journal Supplement Series, 118, 455

\bibitem[{{Sawada} {et~al.}(2004){Sawada}, {Hasegawa}, {Handa}, \&
  {Cohen}}]{Sawada_GC_face_on}
{Sawada}, T., {Hasegawa}, T., {Handa}, T., \& {Cohen}, R.~J. 2004, \mnras, 349,
  1167

\bibitem[{{Sofue}(1995)}]{sofue95}
{Sofue}, Y. 1995, \pasj, 47, 527

\bibitem[{{Tsuboi} {et~al.}(1999){Tsuboi}, {Handa}, \&
  {Ukita}}]{Tsuboi_GC_CSmaps}
{Tsuboi}, M., {Handa}, T., \& {Ukita}, N. 1999, \apjs, 120, 1

\bibitem[{van Eldik(2015)}]{vanEldik201545}
van Eldik, C. 2015, Astroparticle Physics, 71, 45

\bibitem[{{Vovk} {et~al.}(2018){Vovk}, {Strzys}, \& {Fruck}}]{skyprism}
{Vovk}, I., {Strzys}, M., \& {Fruck}, C. 2018, \aap, 619, A7

\bibitem[{Zanin {et~al.}(2013)Zanin, Carmona, Sitarek, Colin, \&
  Frantzen}]{zanin2013mars}
Zanin, R., Carmona, E., Sitarek, J., Colin, P., \& Frantzen, K. 2013, in Proc.
  of the 33st International Cosmic Ray Conference, Rio de Janeiro, Brasil

\end{thebibliography}

\end{document}